\documentclass[aps,prd,preprint,showpacs]{revtex4}
\usepackage{epsfig}
\usepackage{amsmath}
\usepackage[hyperref]{hyperref}

\newcommand{\Pomeron}{I\!\!P}

\begin{document}

\title{The dual parameterization of the proton generalized parton distribution
functions $H$ and $E$ and description of the DVCS cross sections and asymmetries}

\author{V. GUZEY}
\email{vadim.guzey@tp2.rub.de}
\affiliation{Institut f{\"u}r Theoretische Physik II, Ruhr-Universit{\"a}t Bochum,
  D-44780 Bochum, Germany}
\email{vadim.guzey@tp2.rub.de}

\author{T. TECKENTRUP}
\email{tobias.teckentrup@tp2.rub.de}
\affiliation{Institut f{\"u}r Theoretische Physik II, Ruhr-Universit{\"a}t Bochum,
  D-44780 Bochum, Germany}

\pacs{13.60.-r,12.38.Lg}
\preprint{RUB-TP2-04/2006}

\begin{abstract}
 
We develop the minimal model of a new leading order parameterization
of GPDs introduced by Shuvaev and Polyakov.
The model for GPDs $H$ and $E$ is formulated in terms of the forward quark distributions, the Gegenbauer moments of the $D$-term and the forward limit
of the GPD $E$.
The model is designed primarely for small and medium-size values of $x_B$, $x_B \leq 0.2$. 
 We examined two different models of the $t$-dependence of
the GPDs: The factorized exponential model and the non-factorized Regge-motivated
model. Using our model, we successfully described the DVCS cross section measured
by H1 and ZEUS, the moments of the beam-spin $A_{LU}^{\sin \phi}$, 
beam-charge $A_{C}^{\cos \phi}$ and transversely-polarized target $A_{UT}^{\sin \varphi \cos \phi}$ DVCS asymmetries measured by HERMES and 
$A_{LU}^{\sin \phi}$ measured by CLAS.
The data on $A_{C}^{\cos \phi}$ prefers the Regge-motivated model of the 
$t$-dependence of
the GPDs. The data on $A_{UT}^{\sin \varphi \cos \phi}$ indicates that the
$u$ and $d$ quarks carry only a small fraction of the proton total angular momentum.

\end{abstract}

\maketitle

\section{Introduction}
\label{sec:introduction}

Generalized parton distributions (GPDs) parameterize non-perturbative 
parton correlation functions in hadronic 
targets~\cite{Mueller:1998fv,Ji:1998pc,Radyushkin:2000uy,Goeke:2001tz,Belitsky:2001ns,Diehl:2003ny,Belitsky:2005qn}.
The GPDs generalize and interpolate between the common parton distributions and form factors.
Collinear factorization theorems for deeply virtual Compton scattering~\cite{Collins:1998be} 
and for hard electroproduction of mesons~\cite{Collins:1996fb} give a theoretical possibility
 to
experimentally constrain the GPDs. However, since the GPDs are functions of three arguments
(excluding the known dependence on the renormalization scale)  and since experimental 
observables involve convolution of the GPDs with hard scattering coefficients and not
 the GPDs themselves,
it is very difficult to experimentally constrain the GPDs. Therefore, there is 
continuing  interest in modeling GPDs using various 
 models of the hadronic 
structure~\cite{Ji:1997gm,Petrov:1998kf,Tiburzi:2001ta,Tiburzi:2002tq,Scopetta:2003et,Tiburzi:2004mh,Mineo:2005qr,Radyushkin:1998es,Musatov:1999xp,Pobylitsa:2002vw}.

The most convenient and widely used parameterization of GPDs is based on the double 
distribution (DD)
model introduced by Radyushkin~\cite{Radyushkin:1998es,Musatov:1999xp}.
Adding to the DD model the so-called $D$-term~\cite{Polyakov:1999gs}, which is required to
restore the full form of polynomiality of the DD model, 
one obtains a simple, almost analytical,
parameterization of GPDs, which can be readily used for the calculation of various observables,
see e.g.~\cite{Goeke:2001tz}. However, such a parameterization of the GPDs has several
phenomenologically unsatisfactory features. First,  the successful description of the low
Bjorken $x$
HERA data on the total DVCS cross section requires a very specific ($\xi$-independent) 
 shape for 
the input GPDs, which is very different from the input required for the description of the 
DVCS asymmetries measured at higher $x$~\cite{Belitsky:2001ns,Freund:2001hd}.
Second, the parameterization ''does not commute'' with QCD evolution, i.e.~it serves
only to define the input for  QCD evolution of GPDs at a certain initial scale $\mu_0$.
The result of the QCD evolution to the higher scale $\mu$ cannot be generally expressed 
in the form used for the input. Therefore, one has to separately perform the rather 
non-trivial QCD evolution of GPDs.
Third, the parameterization does not allow for an intuitive physical motivation and
interpretation, see~\cite{Freund:2002ff,Polyakov:2002yz} for a discussion of the
 physics of GPDs.

In this paper, we continue and extend the analysis~\cite{Guzey:2005ec} of a new
parameterization of proton GPDs first introduced by Shuvaev and
 Polyakov~\cite{Polyakov:2002wz}.

Initially, the dual representation of quark GPDs of the pion was derived by 
Polyakov~\cite{Polyakov:1998ze} as a formal solution
reproducing  Mellin moments of the pion GPDs. In turn, the moments were related by crossing
to the moments of the two-pion distribution amplitude, which was expanded in terms of 
eigenfunctions of the corresponding QCD evolution equation 
(the Gegenbauer polynomials $C_n^{3/2}$)
 and eigenfunctions of the operator of the relative orbital moment of the pion pair 
(the Legendre polynomials $P_l$). The resulting parameterization of GPDs was termed
dual because the idea of its derivation follows the hypothesis of duality of soft 
hadron-hadron interactions, which is the assumption that
 the $2 \to 2$ scattering amplitude in the $s$-channel 
can be represented as an infinite series over $t$-channel exchanges~\cite{Alfaro}.
In the context of the quark GPDs of the pion, the dual parameterization implies
 that the GPDs are formally given by an infinite
sum of $t$-channel resonances, which build up the two-pion  distribution amplitude.

In the successive work, Shuvaev and Polyakov postulated that a similar dual 
parameterization holds for proton singlet GPDs~\cite{Polyakov:2002wz}
\begin{eqnarray}
&&H^i(x,\xi,t,\mu^2)=\sum_{\substack{n=1\\  {\rm odd}}}^{\infty} \sum_{\substack{l=0 \\ {\rm even}}}^{n+1}
B_{nl}^i(t,\mu^2)\, \theta\left(\xi-|x|\right)\left(1-\frac{x^2}{\xi^2}\right)\,
C_n^{3/2}\left(\frac{x}{\xi}\right)\,P_l\left(\frac{1}{\xi}\right) \,,\nonumber\\
&&E^i(x,\xi,t,\mu^2)=\sum_{\substack{n=1\\  {\rm odd}}}^{\infty} \sum_{\substack{l=0 \\ {\rm even}}}^{n+1}
C_{nl}^i(t,\mu^2)\, \theta\left(\xi-|x|\right)\left(1-\frac{x^2}{\xi^2}\right)\,
C_n^{3/2}\left(\frac{x}{\xi}\right)\,P_l\left(\frac{1}{\xi}\right) \,. 
\label{eq:def}
\end{eqnarray}
In this equation,  the superscript $i$ denotes the quark flavor;
$x$, $\xi$, $t$ and $\mu^2$ stand for the usual arguments of GPDs~\cite{Goeke:2001tz};
$B_{nl}^i$ and $C_{nl}^i$ denote the unknown form factors.
In this work, we restrict ourselves only to the helicity-even GPDs $H$ and $E$ and do not consider the two remaining helicity-even GPDs  $\tilde{H}$ and $\tilde{E}$.
Since we are concerned with DVCS observables, which probe the singlet combinations
of GPDs, we shall consider only singlet GPDs:  
The left hand side of Eq.~(\ref{eq:def}) represents the singlet combinations of
the GPDs, $H^i(x,\xi,t) \equiv H^i(x,\xi,t)-H^i(-x,\xi,t)$ and
 $E^i(x,\xi,t) \equiv E^i(x,\xi,t)-E^i(-x,\xi,t)$, which are antisymmetric with 
respect to $x$.

It should be stressed that the series in Eq.~(\ref{eq:def}) diverge and, hence, 
one should explain how this should be understood. As follows from the
derivation~\cite{Polyakov:1998ze}, Eq.~(\ref{eq:def}) is nothing but a shorthand notation 
for the $x$-moments of the GPDs. Therefore, the formal representation of
Eq.~(\ref{eq:def}) should be understood as a generalized mathematical function,
which acting on the polynomials of $x$, reproduces correctly the 
Mellin moments of the corresponding GPDs. Note that the divergence in Eq.~(\ref{eq:def})
 is analogous to
the divergence of any dual representation of  
the $2 \to 2$ scattering amplitude in soft hadronic physics: 
The $s$-channel series must diverge to reproduce infinities (poles) of the amplitude
in the crossed $t$-channel~\cite{Alfaro}.

Naturally, since the series in Eq.~(\ref{eq:def}) diverge, one cannot use them to study the GPDs as
functions of their variables. However,
since Eq.~(\ref{eq:def}) fixes all Mellin moments of the GPDs, one can 
readily obtain different representations of the GPDs, which would give continuous and
finite  GPDs. For instance, expanding  Eq.~(\ref{eq:def}) over the Gegenbauer
polynomials $C_n^{3/2}(x)$ 
on the interval $-1 \leq x \leq 1$, one obtains~\cite{Belitsky:1997pc}
\begin{equation}
H^i(x,\xi,t,\mu^2)=(1-x^2) \sum_{\substack{n=1\\  {\rm odd}}}^{\infty} A_n^i(\xi,t,\mu^2)C_n^{3/2}(x) \,,
\label{eq:def:conv}
\end{equation}
where
\begin{equation}
A_n^i(\xi,t,\mu^2)=-\frac{2n+3}{(n+1)(n+2)} \sum_{\substack{p=1\\  {\rm odd}}}^{n} \xi R_{np}(\xi)
\frac{(p+1)(p+2)}{2p+3} \sum_{\substack{l=0\\  {\rm even}}}^{p+1} B_{pl}^i(t,\mu^2) P_l\left(\frac{1}{\xi}\right) \,.
\label{eq:an}
\end{equation}
The functions $R_{np}(\xi)$ are polynomials of the order $n$ of the variable  $\xi$,
which are  defined in terms of the hypergeometric function,
\begin{equation}
R_{np}(\xi)=\left(-1\right)^{\frac{n+p}{2}} \frac{\Gamma(\frac{3}{2}+\frac{n+p}{2})}{\Gamma(\frac{n-p}{2}+1)\Gamma(\frac{3}{2}+p)} \xi^p \ _2F_1\left(\frac{p}{2}-\frac{n}{2},\frac{3}{2}+\frac{n}{2}+\frac{p}{2},\frac{5}{2}+p; \xi^2\right) \,.
\label{eq:rnp}
\end{equation}
The representation for the GPDs $E^i$ is obtained by replacing the
form factors $B_{nl}^i$ by $C_{nl}^i$.

The dependence of the GPDs $H^i$ and $E^i$ on the renormalization scale $\mu$ is contained in
the form factors $B_{nl}^i$. By construction~(\ref{eq:def}), $B_{nl}^i$ are proportional 
to the $n$th conformal
moment of the GPDs, which is multiplicatively renormalized to the leading $\alpha_s$ 
accuracy~\cite{Belitsky:1997pc}. 
Therefore, under leading order (LO) QCD evolution, 
the form factors $B_{nl}^i$ have a very 
simple, well-established $\mu^2$-dependence,
\begin{equation}
B_{nl}^i(\mu^2)=B_{nl}^i(\mu_0^2) \, \left(\frac{\ln(\mu_0^2/\Lambda^2)}{\ln(\mu^2/\Lambda^2)}
\right)^{\gamma_n/B} \,,
\label{eq:anom}
\end{equation}
where $B=11-(2/3) N_{{\rm flav}}$ ($N_{{\rm flav}}$ is the number of active parton
flavors);
 $\gamma_n$ are  LO non-singlet anomalous 
dimensions~\cite{Lepage:1979zb,Efremov:1979qk}.
 Alternatively, as will be clear from the
following sections, the $\mu^2$-dependence of $B_{nl}^i$ for all $l$ is given by the  
$\mu^2$-dependence of the $n+1$ Mellin moment of the forward singlet quark distribution,
$\int^{1}_{-1} dx\, x^n q^i(x,\mu^2)$. 

In addition to the simple LO evolution of 
$B_{nl}^i$, the DVCS amplitude has a very simple form in terms of $B_{nl}^i$ also only to the LO
accuracy. Therefore, we shall use the dual parameterization of the GPDs as a LO 
parameterization.

\section{Minimal model of the dual parameterization of GPDs $H^i$ and $E^i$}
\label{sec:model}

In this section, we consider a minimal model of the 
dual parameterization of the GPDs $H^i$ and $E^i$, which 
can be formulated in terms of the forward limit of the GPDs $H^i$ and $E^i$ and 
the Gegenbauer moments of the $D$-term.
The $t$-dependence of the GPDs will be modelled separately.

\subsection{Derivation of the minimal model}

As explained in the Introduction, the dual representation in the form
of Eqs.~(\ref{eq:def:conv}), (\ref{eq:an}) and (\ref{eq:rnp}) gives finite and
continuous expressions for GPDs $H^i$ and $E^i$. However, these equations
are impractical to use since one has to sum an infinite series of large
sign-alternating terms.
An elegant method to sum the series of 
Eq.~(\ref{eq:def:conv}) was offered by Polyakov and Shuvaev~\cite{Polyakov:2002wz}.
The method consists in the introduction of a set of functions whose Mellin moments generate the form factors
$B_{nl}^i$ and $C_{nl}^i$,
\begin{eqnarray}
B_{n \, n+1-k}^i(t,\mu^2)=\int^1_0 dx \,x^n Q_k^i(x,t,\mu^2) \,, \nonumber\\
C_{n \, n+1-k}^i(t,\mu^2)=\int^1_0 dx \,x^n R_k^i(x,t,\mu^2) \,.
\end{eqnarray}
Note that for the singlet combinations of the GPDs that we consider in this paper,
$n$ is odd and $k$ is even.
Using the methods detailed in Appendix~B~\cite{Polyakov:2002wz}, one obtains the following
representation of the GPDs $H^i$ in terms of the generating functions
$Q_k^i$ (the GPDs $E^i$ are obtained by replacing $Q_k^i$ by $R_k^i$)
\begin{eqnarray}
&&H^i(x,\xi,t,\mu^2)=\sum_{\substack{k=0\\  {\rm even}}}^{\infty}\Big[\frac{\xi^k}{2} \left(H^{i\,(k)}(x,\xi,t,\mu^2)-H^{i\,(k)}(-x,\xi,t,\mu^2)\right) \nonumber\\
&+&\left(1-\frac{x^2}{\xi^2}\right)\theta \left(\xi-|x|\right) \sum_{\substack{l=1\\  {\rm odd}}}^{k-3}C^{3/2}_{k-l-2}\left(\frac{x}{\xi}\right) P_l\left(\frac{1}{\xi}\right) \int^1_0 dy\, y^{k-l-2}\, Q_k^i(y,t,\mu^2) \Big] \,,
\label{eq:hq}
\end{eqnarray}
where
\begin{eqnarray}
H^{i\,(k)}(x,\xi,t,\mu^2)&=&\frac{1}{\pi}\int^1_0 \frac{dy}{y}\left[\left(1-y \frac{\partial}{\partial y}\right) Q_k^i(y,t,\mu^2) \right] \int^1_{-1} ds \frac{x_s^{1-k}}{\sqrt{2 x_s-x_s^2-\xi^2}}
\theta(2 x_s-x_s^2-\xi^2) \nonumber\\
&-& \lim_{y \to 0} Q_k^i(y,t,\mu^2) \int^1_{-1} ds \frac{x_s^{1-k}}{\sqrt{2 x_s-x_s^2-\xi^2}}
\theta(2 x_s-x_s^2-\xi^2) \,.
\end{eqnarray}
Note that the last line in the expression for $H^{i\,(k)}$ is missing in the original 
formulation~\cite{Polyakov:2002wz}: Its presence was noticed in Ref.~\cite{RMainz}.

In the present work, we consider a minimal version of the dual representation, which 
consists in retaining only the contributions of the generating functions
$Q_0^i$ and $Q_2^i$ to the GPDs $H^i$ and functions $R_0^i$ and $R_2^i$
 to the GPDs $E^i$.
This means that we keep only the form factors $B^i_{n n+1}$, $B^i_{n n-1}$, 
$C^i_{n n+1}$ and $C^i_{n n-1}$, i.e.~we keep only the maximal $l=n+1$ and
next-to-maximal $l=n-1$ orbital momenta in Eq.~(\ref{eq:def}).
The main motivation for such an approximation is the prefactor $\xi^k$ in Eq.~(\ref{eq:hq}):
In the HERA kinematics ($\xi < 0.005$), the contribution
of $Q_k^i$ and $R_k^i$ with $k \geq 2$ is kinematically suppressed.
In the HERMES kinematics ($\xi < 0.1$),
we keep $Q_2^i$ and  $R_2^i$ as a first correction.
This is also true for the contribution of the second line in Eq.~(\ref{eq:hq}).

The formal representation~(\ref{eq:def}) enables one to readily establish the
connection between the Mellin moments of the GPDs $H^i$ and the form factors $B^i_{nl}$
(similar equation holds for $E^i$)
\begin{equation}
\int^1_ {-1} dx\, x^N \, H^i(x,\xi,t,\mu^2)= \xi^{N+1}\sum_{\substack{n=1\\  {\rm odd}}}^{N}
\sum_{\substack{l=0\\  {\rm even}}}^{n+1} B_{nl}^i(t,\mu^2) P_l\left(\frac{1}{\xi}\right)
\frac{\Gamma\left(\frac{3}{2}\right) \Gamma(N+1) (n+1) (n+2)}{2^N \Gamma\left(\frac{N-n}{2}+1\right) \Gamma\left(\frac{N+n}{2}+\frac{5}{2}\right)} \,.
\label{eq:mm}
\end{equation}
Taking the $\xi \to 0$ limit in this equation, one determines 
the form factors $B_{nn+1}^i$ and $C_{nn+1}^i$, 
\begin{eqnarray}
B_{nn+1}^i(t,\mu^2)&=&\frac{2n +3}{2n+4} \int^{1}_{-1} dx \,x^n H^i(x,0,t,\mu^2) \nonumber\\
& \equiv &
\frac{2n +3}{2n+4} \int^{1}_{0} dx \,x^n \left(q^i(x,t,\mu^2)+\bar{q}^i(x,t,\mu^2)\right)\,,
\nonumber\\
C_{nn+1}^i(t,\mu^2)&=&\frac{2n +2}{2n+4} \int^{1}_{-1} dx \,x^n E^i(x,0,t,\mu^2) \nonumber\\
& \equiv &\frac{2n +3}{2n+4} \int^{1}_{0} dx \,x^n \left(e^i(x,t,\mu^2)+\bar{e}^i(x,t,\mu^2)\right) \,.
\label{eq:normal}
\end{eqnarray}
In the forward limit $t \to 0$,  $q^i(x,t,\mu^2)+\bar{q}^i(x,t,\mu^2)$ become the
singlet combination of  forward quark parton distributions and 
 $e^i(x,t,\mu^2)+\bar{e}^i(x,t,\mu^2)$ become the unknown forward limit of the singlet
 combination of GPDs $E^i$.

Since Eqs.~(\ref{eq:normal}) fix all $B_{nn+1}^i$ and $C_{nn+1}^i$, they
completely determine the generating functions $Q_0^i$ and $R_0^i$~\cite{Polyakov:2002wz,RMainz}
\begin{eqnarray}
Q_0^i(x,t,\mu^2)=q^i(x,t,\mu^2)+\bar{q}^i(x,t,\mu^2)-\frac{x}{2} \int^1_x \frac{dz}{z^2} 
\left(q^i(z,t,\mu^2)+\bar{q}^i(z,t,\mu^2)\right) \,, \nonumber\\
R_0^i(x,t,\mu^2)=e^i(x,t,\mu^2)+\bar{e}^i(x,t,\mu^2)-\frac{x}{2} \int^1_x \frac{dz}{z^2} 
\left(e^i(z,t,\mu^2)+\bar{e}^i(z,t,\mu^2)\right) \,.
\label{eq:q0}
\end{eqnarray}
Therefore, up to the $t$-dependence, the functions $Q_0^i$ and $R_0^i$ are completely constrained
by the forward parton distributions and the forward limit of the GPDs $E^i$.
Note that Eqs.~(\ref{eq:q0}) are valid at all scales $\mu^2$.

Turning to the generating functions $Q_2^i$, we notice that their modeling is
more ambiguous as compared to the functions $Q_0^i$ since only the Mellin moments of 
$Q_2^i$ are constrained. The constraints can be derived as follows. Considering the Mellin moments of 
the GPDs $H^i$, see Eq.~(\ref{eq:mm}), we notice that the coefficient in front of $\xi^{N+1}$, which is
denoted $h_{N+1}^{(N)}$~\cite{Goeke:2001tz}, is
\begin{equation}
h_{N+1}^{i\,(N)}=\sum_{\substack{n=1\\  {\rm odd}}}^{N}
\sum_{\substack{l=0\\  {\rm even}}}^{n+1} B_{nl}^i(t,\mu^2) P_l\left(0\right)
\frac{\Gamma\left(\frac{3}{2}\right) \Gamma(N+1) (n+1) (n+2)}{2^N \Gamma\left(\frac{N-n}{2}+1\right) \Gamma\left(\frac{N+n}{2}+\frac{5}{2}\right)} \,.
\label{eq:hn}
\end{equation}
On the other hand, decomposing the $D$-term in terms of its Gegenbauer moments
\begin{equation}
D^i(z,t,\mu^2)=(1-z^2) \sum_{\substack{n=1\\  {\rm odd}}}^{\infty} d_n^i(t,\mu^2) \,C_n^{3/2}(z)
\label{eq:dterm}
\end{equation}
and using the definition
\begin{equation}
h_{N+1}^{i\,(N)}=\int^1_{-1} dz\,z^N \,D^i(z,t,\mu^2) \,,
\label{eq:hndterm}
\end{equation}
one obtains the desired relation between the $D$-term and the form factors $B^i_{nl}$
\begin{equation}
d_n^i(t,\mu^2) =\sum_{l=0}^{n+1} B_{nl}^i(t,\mu^2) P_l\left(0\right) \,.
\label{eq:dtermbnl}
\end{equation}
 
Within the framework of the minimal version of the dual parameterization, we keep 
only the form factors with $l=n\pm1$ and, hence, obtain 
\begin{equation}
B_{nn-1}^i(t,\mu^2)=\frac{n}{n+1}B_{nn+1}^i(t,\mu^2)+\frac{d_n^i(t,\mu^2)}{P_{n-1}(0)} \,.
\label{eq:bn}
\end{equation}
This equation allows us to constrain the generating function $Q_2^i$ as follows. 
Decomposing $Q_2^i$ as
\begin{equation}
Q_2^i(x,t,\mu^2)=Q_0^i(x,t,\mu^2)-\int^1_x \frac{dz}{z}\, Q_0^i(z,t,\mu^2)+\tilde{Q}_{2}^i(x,t,\mu^2)
\label{eq:q2tilde}
\end{equation} 
and substituting this in Eq.~(\ref{eq:bn}), one obtains the following constraint
 on the new unknown function $\tilde{Q}_{2}^i$
\begin{equation}
\int^1_0 dx \,x^n \, \tilde{Q}_{2}^i(x,t,\mu^2)=\frac{d_n^i(t,\mu^2)}{P_{n-1}(0)} \,.
\label{eq:q2tildemoment}
\end{equation} 

Turning to the generating functions $R_2^i$ for the GPDs $E^i$, we notice that
since the $D$-term contribution to the GPDs $E^i$ and $H^i$ are equal and opposite
in sign, see e.g.~\cite{Goeke:2001tz}, then
\begin{equation}
C_{nn-1}^i(t,\mu^2)=\frac{n}{n+1}C_{nn+1}^i(t,\mu^2)-\frac{d_n^i(t,\mu^2)}{P_{n-1}(0)} \,.
\label{eq:cn}
\end{equation}
Therefore, the functions $R_2^i$ can be written in the form 
\begin{equation}
R_2^i(x,t,\mu^2)=R_0^i(x,t,\mu^2)-\int^1_x \frac{dz}{z}\, R_0^i(z,t,\mu^2)-\tilde{Q}_{2}^i(x,t,\mu^2) \,.
\label{eq:r2tilde}
\end{equation} 
Note that this representation for $R_2^i$ was first suggested by
Polyakov and Shuvaev, see Eq.~(56) of~\cite{Polyakov:2002wz}.

\subsection{Details of the minimal model}

So far we have presented rather general consideration involved in the construction
of the minimal model of the dual parameterization of the GPDs $H^i$ and $E^i$. 
In the following, we shall discuss such details of the parameterization as
the modeling of the functions $\tilde{Q}_2^i$  and  $e^i+\bar{e}^i$ and the 
modeling of the $t$-dependence of $Q_k^i$ and $R_k^i$.

In the following discussion of $\tilde{Q}_2^i$ and $e^i+\bar{e}^i$, we assume that
$t=0$ and do not explicitly write the $t$-dependence.
The shape of the functions $\tilde{Q}_2^i$ is unconstrained.
We assume a simple form for $\tilde{Q}_{2}^i$,
\begin{equation}
\tilde{Q}_{2}^i(x,\mu^2)=(1-x)\left(a^i(\mu^2)+b^i(\mu^2)\,x +c^i(\mu^2)\,x^2\right) \,,
\label{eq:simpleform}
\end{equation} 
with the coefficients $a^i$, $b^i$ and $c^i$ fixed by the constraint of 
Eq.~(\ref{eq:q2tildemoment}) evaluated for $n=1$, 3 and 5.
Note that the $\mu^2$-dependence of $d_n^i$ and, hence, the $\mu^2$-dependence
of $a^i$, $b^i$ and $c^i$ is given by Eq.~(\ref{eq:anom}).

The singlet combination of the first three Gegenbauer moments of the $D$-term was evaluated in the 
chiral quark soliton model at the low normalization point $\mu_0 \approx 0.6$ GeV~\cite{Kivel:2000fg}
\begin{eqnarray}
&& d_n^u(\mu_0^2)+d_n^d(\mu_0^2)=R_n \left[(M_2^u(\mu_0^2)+M_2^d(\mu_0^2)\right]=R_n \,,
 \nonumber\\
&& R_1=-8\,, \quad R_3=-2.4\,, \quad R_5=-0.8 \,,
\label{eq:soliton}
\end{eqnarray}
where 
$M_2^i$ is the proton momentum fraction carried by the quark and antiquark of
 the flavor~$i$.
 Note that the last equality in the first line of  Eq.~(\ref{eq:soliton})
comes from the fact that at the low normalization point $\mu_0$, in the SU(2)-symmetric
chiral quark soliton model, the $u$ and $d$ quarks carry the entire proton momentum.
At higher $\mu^2$ of the order a few GeV$^2$, the quarks carry about half of the proton momentum,
 which
reduces the numerical values in Eq.~(\ref{eq:soliton}) by the factor two~\cite{Kivel:2000fg}. 
In our analysis, we use
\begin{equation}
d_n^u(\mu_0^2)=R_n \,M_2^u(\mu_0^2) \,, \quad d_n^d(\mu_0^2)=R_n \,M_2^d(\mu_0^2) \,, \quad d_n^s=0 \,.
\label{eq:soliton2}
\end{equation}
The quark momentum fractions are evaluated using the leading order CTEQ5L parton
distributions~\cite{Lai:1999wy}.

Since the GPDs $E^i$ decouple in the forward limit, the functions $e^i+\bar{e}^i$
in Eq.~(\ref{eq:normal}) are unconstrained. In this work, we follow the simple model
of Ref.~\cite{Goeke:2001tz},
which correctly reproduces the forward limit of the first moment of the GPDs $E^i$ and which 
allows to vary the fraction of the nucleon spin carried by quarks
\begin{eqnarray}
&&e^i(x,\mu^2)= A_i(\mu^2)\, q^i_{{\rm val}}(x,\mu^2)+\frac{B_i(\mu^2)}{2}\, \delta(x) \,, \nonumber\\
&&\bar{e}^i(x,\mu^2)=\frac{B_i(\mu^2)}{2}\, \delta(x) \,,
\label{eq:emodel}
\end{eqnarray}
where $q^i_{{\rm val}}(x,\mu^2)=q^i(x,\mu^2)-{\bar q}^i(x,\mu^2)$ is the 
valence quark distribution.
Note that unlike $\bar{q}^i$, $\bar{e}^i$ is a symmetric function of $x$.
The coefficients $A_i$ and $B_i$ for $u$ and $d$ quarks are found from the first and second
Mellin moments of the GPDs $E^i$~\cite{Goeke:2001tz}
\begin{eqnarray}
&&A_i(\mu^2)=\frac{2J^i(\mu^2)-M_2^i(\mu^2)}{M_2^{i,{\rm val}}} \,, \nonumber\\
&&B_u(\mu^2)=k_u-2\,A_u(\mu^2) \,, \quad B_d(\mu^2)=k_d-A_d(\mu^2) \,,
\label{eq:largeAB}
\end{eqnarray}
where $J^i$ is the contribution to the proton total angular momentum of the quark with
flavor~$i$; $M_2^{i,{\rm val}}$ is the
proton momentum fraction carried by the valence part of the quark distribution function;
$k_u=1.673$ and $k_d=-2.033$ are quark anomalous magnetic moments.
We assume that for the strange and charm quarks, $e^i(x)=\bar{e}^i(x)=0$.
Note also that the $\mu^2$-dependence of $J^i$ is the same as for $M_2^i$, 
see e.g.~\cite{Goeke:2001tz}.

In summary, the $x$ and $\xi$-dependence of the GPDs $H^i$ is specified by 
Eqs.~(\ref{eq:q0}), (\ref{eq:q2tilde}), (\ref{eq:q2tildemoment}) and (\ref{eq:simpleform}). 
The GPDs $E^i$ are specified by Eqs.~(\ref{eq:q0}), (\ref{eq:r2tilde}), (\ref{eq:emodel})
and (\ref{eq:largeAB}).

The $t$-dependence of the form factors $B^i_{nl}$ and $C^i_{nl}$ has to be specified extra.
In this work, we consider two models of the $t$-dependence. The first model assumes that
all $B^i_{nl}$ and $C^i_{nl}$ and, hence, $H^i$ and $E^i$ have a factorized exponential 
$t$-dependence with the $\mu^2$-dependent slope~\cite{Guzey:2005ec}
\begin{eqnarray}
&&H^i(x,\xi,t,\mu^2)=\exp\left(\frac{B(\mu^2)\,t}{2}\right) H^i(x,\xi,t=0,\mu^2) \,, \nonumber\\
&&E^i(x,\xi,t,\mu^2)= \exp\left(\frac{B(\mu^2)\,t}{2}\right) E^i(x,\xi,t=0,\mu^2) \,,
\label{eq:model1}
\end{eqnarray}
where 
\begin{equation}
B(\mu^2)=7.6 \, \left(1-0.15 \,\ln(\mu^2/2) \right) \ {\rm GeV}^2 \,.
\label{eq:B}
\end{equation}
The decrease of the slope with increasing $\mu^2$ was taken from the model~\cite{Freund:2002qf}.
The normalization of the slope was chosen in order to reproduce the result of
the exponential fit to
the $t$-dependence of the differential DVCS cross section measured by the H1 collaboration
at HERA for $0.1 \leq |t| \leq 0.8$ GeV$^2$ and at $\mu^2=8$ GeV$^2$, 
$B(\mu^2=8 \ {\rm GeV}^2)=6.02  \pm 0.35 \pm 0.39$ GeV$^{-2}$~\cite{Aktas:2005ty}.
Note that a factorized model of the $t$-dependence with the $\mu^2$-independent
slope  commutes with the QCD evolution. While this is not so in our 
case~(\ref{eq:model1}), effects of the $\mu^2$-dependence of the slope
$B(\mu^2)$ on the QCD evolution are numerically small and, thus, have been neglected.

The second model of the $t$-dependence is much more involved: 
It is non-factorized and,
hence, the $t$-dependence non-trivially changes with QCD evolution.
Since the dual parameterization of GPDs is constructed as an
infinite series of $t$-channel exchanges, which resembles
the representation of hadron-hadron scattering amplitudes in Regge theory,
this analogy can serve as a guide for the $t$-dependence of the GPDs. 
In particular, we use the following Regge theory motivated model for 
$q^i(t)$ and $\bar{q}^i(t)$~\cite{Goeke:2001tz}
\begin{eqnarray}
&& q^i(x,t,\mu_0^2)-\bar{q}^i(x,t,\mu_0^2)=q_{{\rm val}}^i(x,t,\mu_0^2)=\left(\frac{1}{x^{\alpha^{\prime}_{{\rm val}}t}}\right) q_{{\rm val}}^i(x,\mu_0^2) \,, \nonumber\\
&& q^i(x,t,\mu_0^2)+\bar{q}^i(x,t,\mu_0^2)=\left(\frac{1}{x^{\alpha^{\prime} t}}\right) \ \left[q^i(x,\mu_0^2)+\bar{q}^i(x,\mu_0^2)\right] \,, \nonumber\\
&&g(x,t,\mu_0^2)=\left(\frac{1}{x^{\alpha_g^{\prime} t}}\right) g(x,\mu_0^2) \,,
\label{eq:regge}
\end{eqnarray}
where $q^i(x,\mu^2)$ and $\bar{q}^i(x,\mu^2)$ are quark and antiquark
forward parton distributions and $g(x,\mu^2)$ is the gluon forward distribution.

The model of Eq.~(\ref{eq:regge}) is specified at some low normalization point. 
 In this work,
$\mu_0^2=1$ GeV$^2$. The functions $q^i(t)$ and $\bar{q}^i(t)$ at 
higher $\mu^2 > \mu_0^2$ are obtained by LO DGLAP evolution at fixed $t$.
While the gluon function $g(x,t,\mu^2)$ does not enter any LO expressions for 
DVCS observables (the hand-bag approximation), one still needs to specify $g(x,t,\mu_0^2)$
for the QCD evolution.

The parameters $\alpha^{\prime}_{{\rm val}}$, $\alpha^{\prime}$ and $\alpha_g^{\prime}$
can be thought of as effective slopes of the corresponding Regge trajectories. 
For the valence quarks,
we use  $\alpha^{\prime}_{{\rm val}}=1.1 (1-x)$ GeV$^{-2}$~\cite{Guidal:2004nd}, which gives
a good description of the nucleon elastic form factors. Numerically similar
options for $\alpha^{\prime}_{{\rm val}}$ were also 
considered in the literature~\cite{Goeke:2001tz,Vanderhaeghen:2002ax,Diehl:2004cx,Diehl:2005wq}.
All of them correspond to $\alpha^{\prime}_{{\rm val}} \approx 0.9-1.0$ GeV$^{-2}$, which is
the typical slope of all known meson and baryon Regge trajectories. 

Drawing an analogy between the parameters $\alpha^{\prime}$ and $\alpha_g^{\prime}$
and the slope of the Pomeron trajectory $\alpha^{\prime}_{\Pomeron}$,
 one would expect that $\alpha^{\prime} \approx \alpha_g^{\prime} \approx \alpha^{\prime}_{\Pomeron}=0.25$ GeV$^{-2}$. However, our analysis of the DVCS cross 
section in Sect.~\ref{sec:cross_section} shows that larger values should be taken.
In this work, we use
\begin{equation}
\alpha^{\prime}=0.9 \  {\rm GeV}^{-2} \,, \quad \alpha_g^{\prime}=0.5 \  {\rm GeV}^{-2} \,.
\label{eq:regge2}
\end{equation}
The inconsistency between the phenomenologically large values of 
$\alpha^{\prime}$ and $\alpha_g^{\prime}$ and the ones expected on the basis of the
Regge theory was discussed in Ref.~\cite{Diehl:2005wq}.

Since the $D$-term does not have a partonic interpretation, we cannot use the 
model of Eq.~(\ref{eq:regge}) to constrain the $t$-dependence of the Gegenbauer moments
$d_n^i$, and, hence, the $t$-dependence of ${\tilde Q}_2^i$.
 Instead, we imploy the results of the lattice calculations
of the $t$-dependence of the first moment of GPDs,
 which were fitted to  the dipole form~\cite{Gockeler:2003jf}, and use
\begin{equation}
d_n^{u,d}(t)= d_n^{u,d}(t=0) \frac{1}{(1-t/M_D^2)^2} \,,
\end{equation}
where $M_D=1.11 \pm 0.20$ GeV in the continuum limit~\cite{Gockeler:2003jf}.

Finally, the same dipole form of the $t$-dependence was assumed for the
$\delta$-function contribution to the functions $e^i$ and $\bar{e}^i$, 
see Eq.~(\ref{eq:emodel}).

\subsection{Predictions for the GPDs $H^i$ and $E^i$}

\begin{figure}[t]
\begin{center}
\epsfig{file=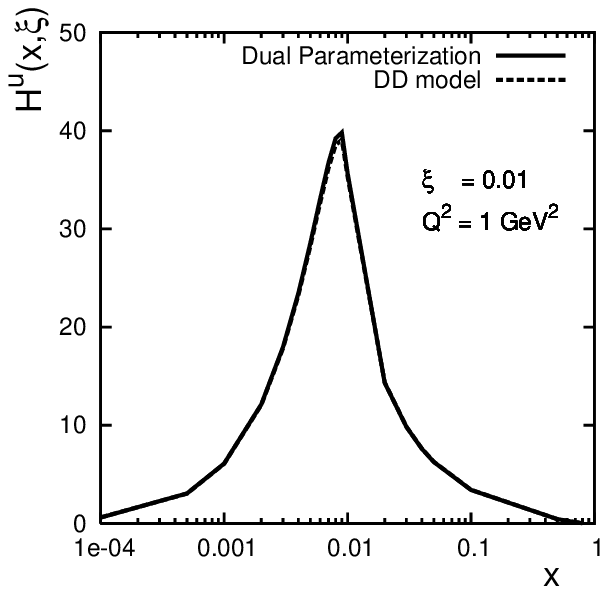,width=8cm,height=8cm}
\epsfig{file=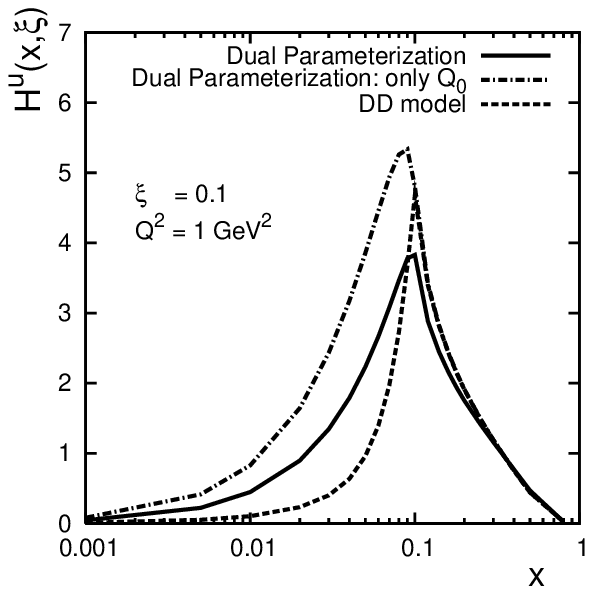,width=8cm,height=8cm}
\epsfig{file=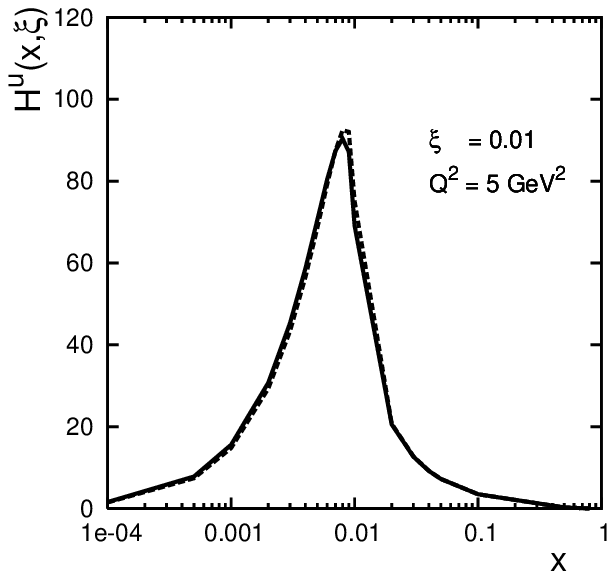,width=8cm,height=8cm}
\epsfig{file=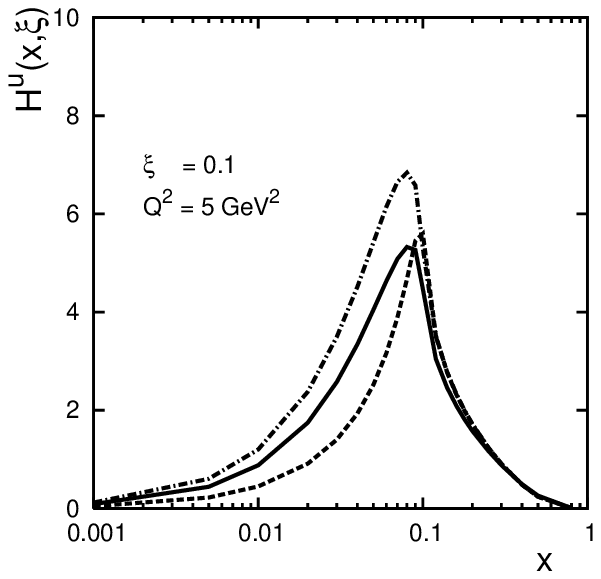,width=8cm,height=8cm}
\caption{The singlet GPD $H^u$ as a function of $x$, $\xi$ and $\mu^2$. 
The dual parameterization results~(\ref{eq:hq}) [solid curves] are compared to 
the predictions of the  DD model~(\ref{eq:radyushkin}) [dashed curves].
The dot-dashed curves correspond to the dual parameterization, when
only the function $Q_0^i$ is retained.
}
\label{fig:Hu}
\end{center}
\end{figure}

\begin{figure}[t]
\begin{center}
\epsfig{file=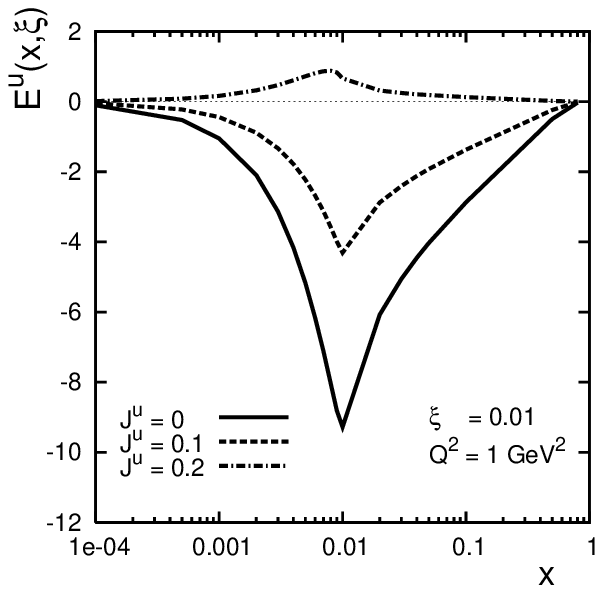,width=8cm,height=8cm}
\epsfig{file=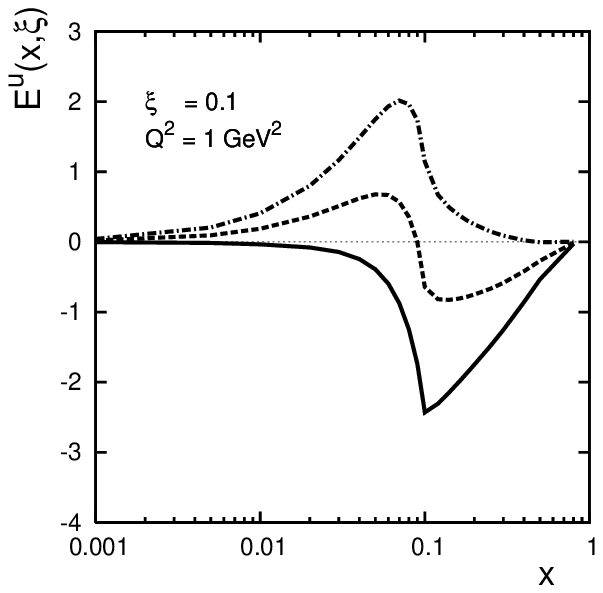,width=8cm,height=8cm}
\epsfig{file=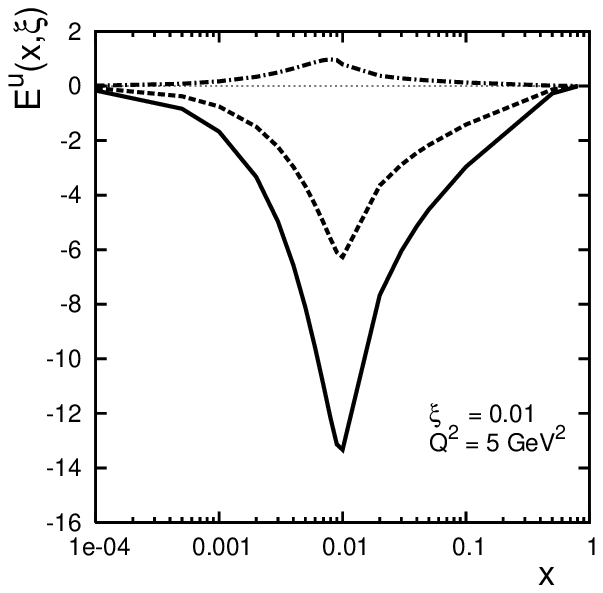,width=8cm,height=8cm}
\epsfig{file=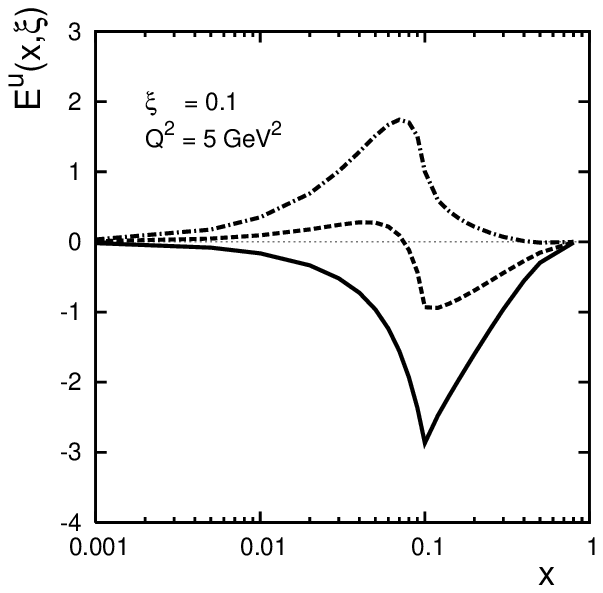,width=8cm,height=8cm}
\caption{The singlet GPD $E^u$ as a function of $x$, $\xi$ and $\mu^2$.
The curves are obtained using the dual parameterization summarized in 
Eqs.~(\ref{eq:hq}), (\ref{eq:r2tilde}), (\ref{eq:emodel}) and (\ref{eq:largeAB}).
The solid curves correspond to $J^u=0$; the dashed curves correspond to $J^u=0.1$;
the dot-dashed curves correspond to $J^u=0.2$.}
\label{fig:Eu}
\end{center}
\end{figure}

\begin{figure}[h]
\begin{center}
\epsfig{file=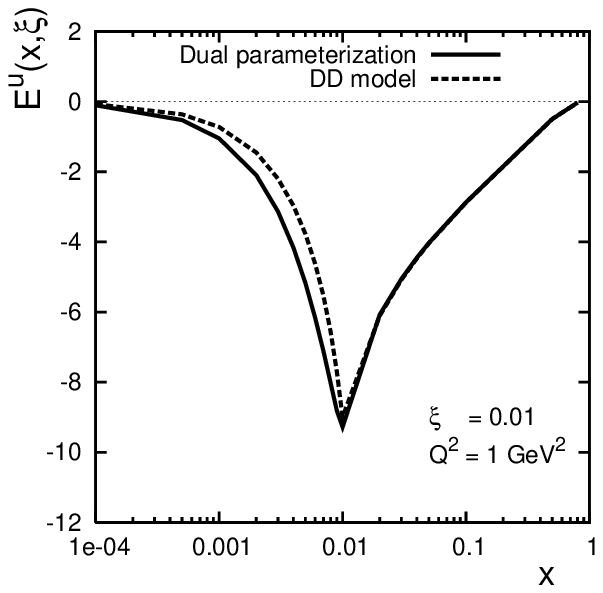,width=8cm,height=8cm}
\epsfig{file=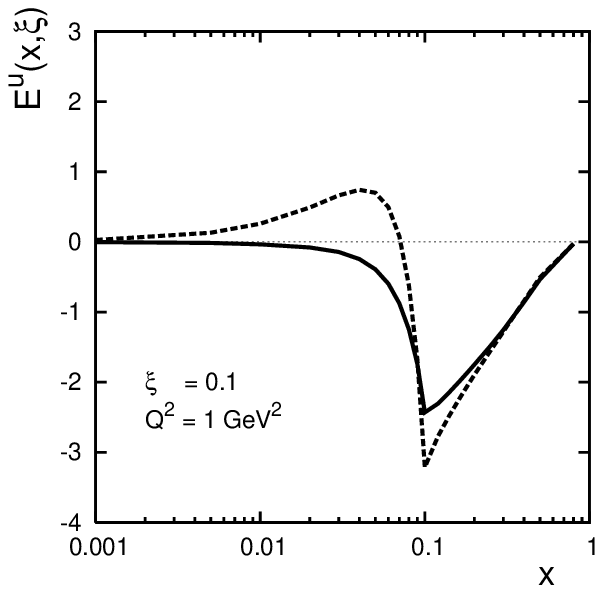,width=8cm,height=8cm}
\caption{The singlet GPD $E^u$ with $J^u=0$. The dual parameterization calculation
(solid curves) is confronted with the DD model calculation (dashed curves).}
\label{fig:Erad}
\end{center}
\end{figure}

The numerical predictions for the $x$, $\xi$ and $\mu^2$-dependence of the GPDs
$H^u$ and $E^u$ at $t=0$ are presented in Figs.~\ref{fig:Hu}, \ref{fig:Eu}
and \ref{fig:Erad} (see the captions).

In Fig.~\ref{fig:Hu}, the results of the dual parameterization are presented as solid
curves. For comparison, the predictions of the double distribution (DD) model
with the $D$-term added are given by dashed curves. In addition, for the $\xi=0.1$ case,
the dot-dashed curves present the calculation using the dual parameterization, when
the contribution of $Q_2^i$ is omitted. Therefore, the deviation of the dot-dashed from
the solid curves can serve as an estimate of the theoretical uncertainty associated with
the modeling of the function $Q_2^i$. Note that for  $\xi=0.01$, the dot-dashed and solid 
curves are indistinguishable: Only the contribution of $Q_0^i$ is important at sufficiently
low $\xi$.

The predictions of the DD model (dashed curves)  for the singlet combination of the GPDs
$H^i$ were made using the standard 
expressions~\cite{Goeke:2001tz,Radyushkin:1998es,Musatov:1999xp} 
\begin{eqnarray}
&&H^i_{{\rm DD}}(x,\xi,\mu^2) \equiv \frac{H^i(x,\xi,\mu^2)-H^i(-x,\xi,\mu^2)}{2} \nonumber\\
&&=
\int^1_{-1} d \beta \int^{1-|\beta|}_{-1+|\beta|} d \alpha \,[\delta(x-\beta-\alpha \xi)-\delta(-x-\beta-\alpha \xi)]
\,h(\beta,\alpha)\,\frac{q^i(\beta,\mu^2)}{2} \nonumber\\
&&+\theta(\xi-|x|) D^i\left(\frac{x}{\xi},\mu^2 \right) \nonumber\\
&&= \int^1_{0} d \beta \int^{1-\beta}_{-1+\beta} d \alpha \left[\delta(x-\beta-\alpha \xi)
-\delta(x+\beta-\alpha \xi) \right]
h(\beta,\alpha) \left(\frac{q^i(\beta,\mu^2)+\bar{q}^i(\beta,\mu^2)}{2} \right) \nonumber\\
&&+\theta(\xi-|x|) D^i \left(\frac{x}{\xi},\mu^2\right)  \,,
\label{eq:radyushkin}
\end{eqnarray}
where 
\begin{equation}
h(\beta,\alpha)=\frac{3}{4} \frac{(1-|\beta|)^2-\alpha^2}{(1-|\beta|)^3} \,,
\end{equation} 
and
\begin{equation}
D^i(z,\mu^2)=\left(1-z^2\right) \sum_{n=1}^5 d_n^i(\mu^2) C_n^{3/2}\left(z\right) \,.
\end{equation}
Note the coefficient $1/2$ in the first line of Eq.~(\ref{eq:radyushkin}): It is 
required to have the correct normalization of the first moment of $H^i_{{\rm DD}}$.

The dual parameterization predictions for the singlet GPD $E^u$ are summarized in
Fig.~\ref{fig:Eu}.
We would like to emphasize that the shape of the GPDs $E^i$
is unknown. The model for the forward limit
of the GPDs $E^i$ that we used, see Eqs.~(\ref{eq:emodel}) and (\ref{eq:largeAB}), was chosen
rather arbitrary and, eventually, might turn out to be wrong. In this model, the
shape of $E^i$ is correlated with the fraction of the proton
total orbital momentum carried by the quark, $J^i$. 
In Fig.~\ref{fig:Eu}, the solid curves correspond to $J^u=0$; the dashed curves correspond 
to $J^u=0.1$; the dot-dashed curves correspond to $J^u=0.2$.

For an alternative model of GPDs $E^i$, we refer the reader to the calculations within
the framework of the chiral quark soliton model~\cite{Ossmann:2004bp}.

In Fig.~\ref{fig:Erad}, the dual parameterization calculation of the singlet $E^u$ with
$J^u=0$ (solid curves) is compared to the DD model calculation (dashed curves). The latter was
performed using Eq.~(\ref{eq:radyushkin}) after the replacement of $q^i$ by $e^i$ and
after changing the sign in front of the $D$-term.

\section{DVCS cross section}
\label{sec:cross_section}

The differential cross section of deeply virtual Compton scattering (DVCS) 
reads~\cite{Goeke:2001tz}
\begin{equation}
\frac{d \sigma}{dQ^2 d x_B dt d \phi}=\frac{1}{32 \,(2 \pi)^4}\, \frac{x_B \,y^2}{Q^4} \,\frac{e^6}{\sqrt{1+ 4m_N^2 x_B^2/Q^2}} |{\bar {\cal T}}_{{\rm DVCS}}|^2 \,,
\label{eq:dvcs}
\end{equation}
where $x_B$, $Q^2$ and  $y=Q^2/(x_B s)$ ($\sqrt{s}$ is the lepton-proton invariant mass)
 are the usual Bjorken variables; 
 $\phi$ is the angle between the plane 
formed by the leptons and the plane formed by the final photon and the final 
proton~\cite{Belitsky:2001ns};
${\cal T}_{{\rm DVCS}}$ is the full DVCS amplitude.
The bar over the DVCS amplitude squared means that we have summed over the
final polarization and averaged over the initial polarizations of
all involved particles.

The results of high-energy DVCS measurements at HERA are usually 
presented in terms of the DVCS cross section on the photon level~\cite{Aktas:2005ty,Adloff:2001cn,Chekanov:2003ya}
\begin{equation}
\sigma_{{\rm DVCS}}(x_B,Q^2)=\frac{1}{\Gamma} \left(\frac{x_B}{y}\right) \int dt\, d \phi \,
\frac{d \sigma}{dQ^2 d x_B dt d \phi} \,,
\end{equation}
where 
\begin{equation}
\Gamma=\frac{\alpha_{{\rm em}} \left(1-y+y^2/2\right)}{\pi Q^2 y} 
\end{equation}
is the flux of the equivalent photons~\cite{Aktas:2005ty}.
Squaring the full DVCS amplitude, averaging over initial polarization and summing
over final polarizations, one obtains the following
unpolarized, $t$-integrated DVCS cross section on the photon level
\begin{equation}
\sigma_{{\rm DVCS}}(x_B,Q^2)=\frac{\pi \alpha_{{\rm em}}^2 x_B^2}{Q^4 \sqrt{1+ 4m_N^2 x_B^2/Q^2}}
\int^{t_{{\rm max}}}_{t_{{\rm min}}} dt \,|{\cal A}_{{\rm DVCS}}(\xi,t,Q^2)|^2 \,,
\label{eq:sigmadvcs}
\end{equation}
where 
\begin{equation}
|{\cal A}_{{\rm DVCS}}(\xi,t,Q^2)|^2=|{\cal H}|^2(1-\xi^2)-
\xi^2 \left({\cal H}^{\ast} {\cal E}+
{\cal H} {\cal E}^{\ast}\right)-|{\cal E}|^2\left(\frac{t}{4 m_N^2}+\xi^2 \right)
\label{eq:Advcs}
\end{equation}
and
\begin{eqnarray}
{\cal H}(\xi,t,Q^2)&=&\sum_i e_i^2 \int^{1}_{0} dx\, H^i(x,\xi,t,Q^2)\left(\frac{1}{x-\xi+i0}+
\frac{1}{x+\xi-i0} \right) \,, \nonumber\\
{\cal E}(\xi,t,Q^2)&=&\sum_i e_i^2 \int^{1}_{0} dx\, E^i(x,\xi,t,Q^2)\left(\frac{1}{x-\xi+i0}+
\frac{1}{x+\xi-i0} \right) \,.
\label{eq:convolution}
\end{eqnarray}
Throughout this paper, the skewedness parameter $\xi$ is related to the Bjorken
variable $x_B$ as $\xi=x_B/(2-x_B)$~\cite{Goeke:2001tz}.
Equations~(\ref{eq:sigmadvcs}) and (\ref{eq:Advcs}) for the unpolarized
DVCS cross section can be also obtained from more general 
expressions derived in Ref.~\cite{Belitsky:2001ns}.

It is important to note that in Eq.~(\ref{eq:convolution}) we used the notation of
Ref.~\cite{Goeke:2001tz}, which differs from the notation of Ref.~\cite{Belitsky:2001ns}
by an overall minus sign. While this is immaterial for the DVCS cross section, 
this matters for the DVCS asymmetries.

One appealing feature of the dual parameterization of GPDs is that the convolution
integrals in Eq.~(\ref{eq:convolution}) can be readily taken and expressed in terms
of the generating functions $Q_n^i$ and $R_n^i$~\cite{Polyakov:2002wz}
\begin{eqnarray}
&&{\cal H}(\xi,t,Q^2)=-\sum_i e_i^2 \int^{1}_{0} \frac{dx}{x}\sum_{k=0}^{\infty}x^k Q_k^i(x,t,Q^2) \left(\frac{1}{\sqrt{1-\frac{2x}{\xi}+x^2}}+\frac{1}{\sqrt{1+\frac{2x}{\xi}+x^2}}-2\delta_{k0}\right) \,, \nonumber\\
&&{\cal E}(\xi,t,Q^2)= \nonumber\\
&&-\sum_i e_i^2 \int^{1}_{0} \frac{dx}{x}\sum_{k=0}^{\infty}x^k R_k^i(x,t,Q^2) \left(\frac{1}{\sqrt{1-\frac{2x}{\xi}+x^2}}+\frac{1}{\sqrt{1+\frac{2x}{\xi}+x^2}}-2\delta_{k0}\right) \,.
\label{eq:Hcal}
\end{eqnarray}

The high-energy HERA data on the total DVCS cross section corresponds to very small 
$\xi$, $\xi < 0.005$, and to small $t$, $t < 1$ GeV$^2$. Therefore,
the contribution of the GPD $E$ to the DVCS cross section is negligible. Moreover,
as discussed in Sect.~\ref{sec:model}, at small $\xi$, the contribution of
$Q^i_n$ with $n \geq 2$ can be safely neglected. Therefore, our predictions for
$\sigma_{{\rm DVCS}}$ within the framework of the dual parameterization are made 
keeping only the functions $Q^i_0$, which, up to the $t$-dependence, are given by the
forward quark distributions.

\begin{figure}[t]
\begin{center}
\epsfig{file=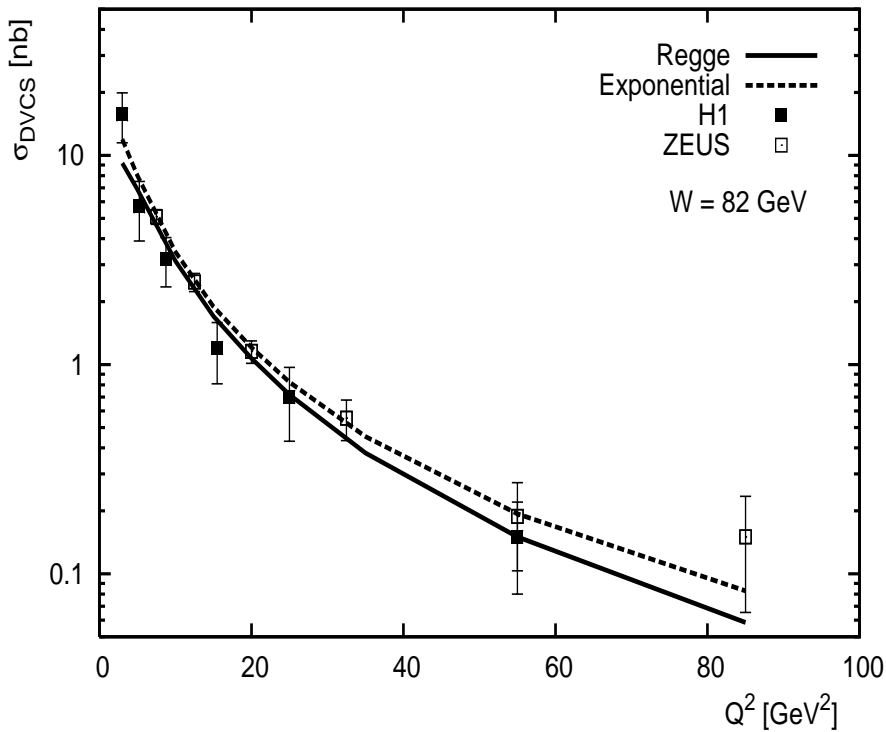,width=12cm,height=10cm}
\caption{The total DVCS cross section $\sigma_{{\rm DVCS}}$ as a function
of $Q^2$ at fixed $W$. The dual parameterization predictions are given by the solid
(Regge model for the $t$-dependence) and dashed (factorized exponential  $t$-dependence)
curves. The experimental points are H1~\protect\cite{Aktas:2005ty}
 and ZEUS~\protect\cite{Chekanov:2003ya}. The statistical and systematic experimental
errors are added in quadrature.
}
\label{fig:q2vadim}
\end{center}
\end{figure}

\begin{figure}[t]
\begin{center}
\epsfig{file=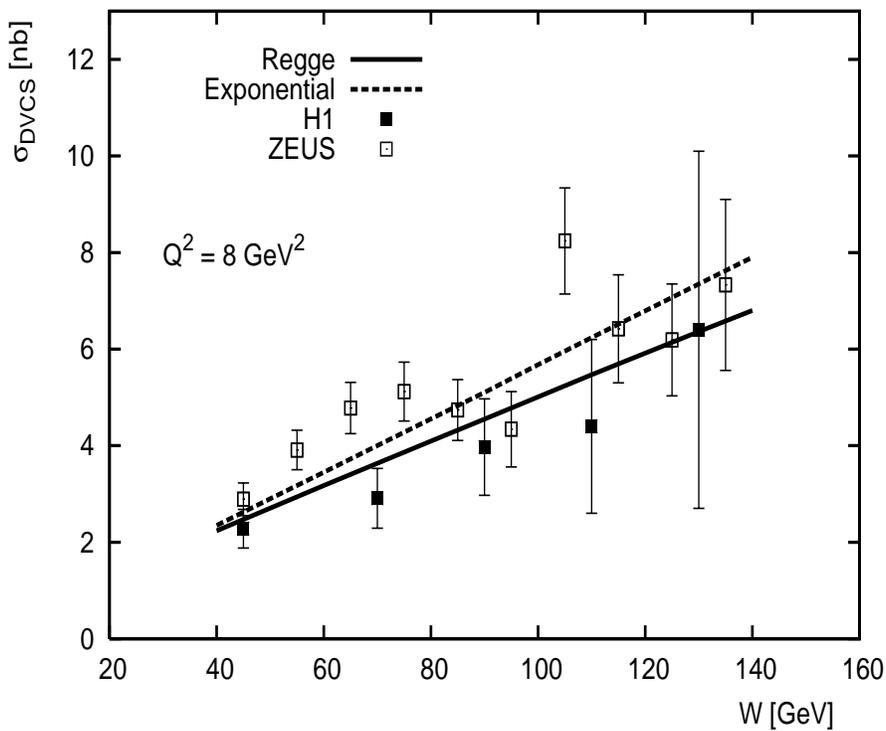,width=12cm,height=10cm}
\caption{The total DVCS cross section $\sigma_{{\rm DVCS}}$ as a function
of $W$ at fixed $Q^2$. The captions are the same as in Fig.~\ref{fig:q2vadim}.}
\label{fig:wvadim}
\end{center}
\end{figure}

Figures~\ref{fig:q2vadim} and  \ref{fig:wvadim} present our predictions
for the $Q^2$ and $W$ dependence of 
$\sigma_{{\rm DVCS}}$. The calculations are performed using the Regge-motivated
$t$-dependence~(\ref{eq:regge}) [solid curves] and the factorized exponential 
$t$-dependence~(\ref{eq:model1}) [dashed curves].
The theoretical predictions are compared to the H1~\cite{Aktas:2005ty}
 and ZEUS~\cite{Chekanov:2003ya} data.

Note that the ZEUS data, which were taken at $W=89$ GeV and $Q^2=9.6$ GeV$^2$, 
have been rescaled to the H1 values of $W=82$ GeV and $Q^2=8$ GeV$^2$
using the fitted $W$ and $Q^2$ dependence of the DVCS cross section measured
by ZEUS, $\sigma_{{\rm DVCS}} \propto W^{0.75}$ and 
$\sigma_{{\rm DVCS}} \propto 1/(Q^2)^{1.54}$~\cite{Chekanov:2003ya}.

One can see from Fig.~\ref{fig:q2vadim} that the absolute value and the
 $Q^2$ dependence of the total DVCS
cross section is
reproduced well using both the non-factorized Regge-motivated~(\ref{eq:regge})
 and factorized exponential~(\ref{eq:model1})
models of the $t$-dependence.
However, at the highest values of $Q^2$, the exponential model of the $t$-dependence
gives somewhat larger $\sigma_{{\rm DVCS}}$ because of the $Q^2$-dependent slope 
$B$~(\ref{eq:B}), which provides a better agreement with the highest $Q^2$ ZEUS point.

Note that the parameters $\alpha^{\prime}$ and $\alpha^{\prime}_g$ of the Regge-motivated
model of the $t$-dependence, see Eq.~(\ref{eq:regge2}),
 were chosen such that the theoretical calculations
reproduce well the absolute value of $\sigma_{{\rm DVCS}}$ in Fig.~\ref{fig:q2vadim}.
Smaller values of $\alpha^{\prime}$ and $\alpha^{\prime}_g$, which would
be closer to $\alpha^{\prime}_{\Pomeron}$, would give inconsistently
large values of $\sigma_{{\rm DVCS}}$.

From Fig.~\ref{fig:wvadim} one can see that the absolute value and the $W$ dependence of 
$\sigma_{{\rm DVCS}}$ is also reproduced well.  The H1 data~\cite{Aktas:2005ty}
somewhat prefers the results of the calculation using the Regge-motivated $t$-dependence.
However, large experimental errors 
at large values of $W$ and a slight discrepancy between the H1 and ZEUS data do
not allow one to draw a more quantitative conclusion.

\begin{figure}[h]
\begin{center}
\epsfig{file=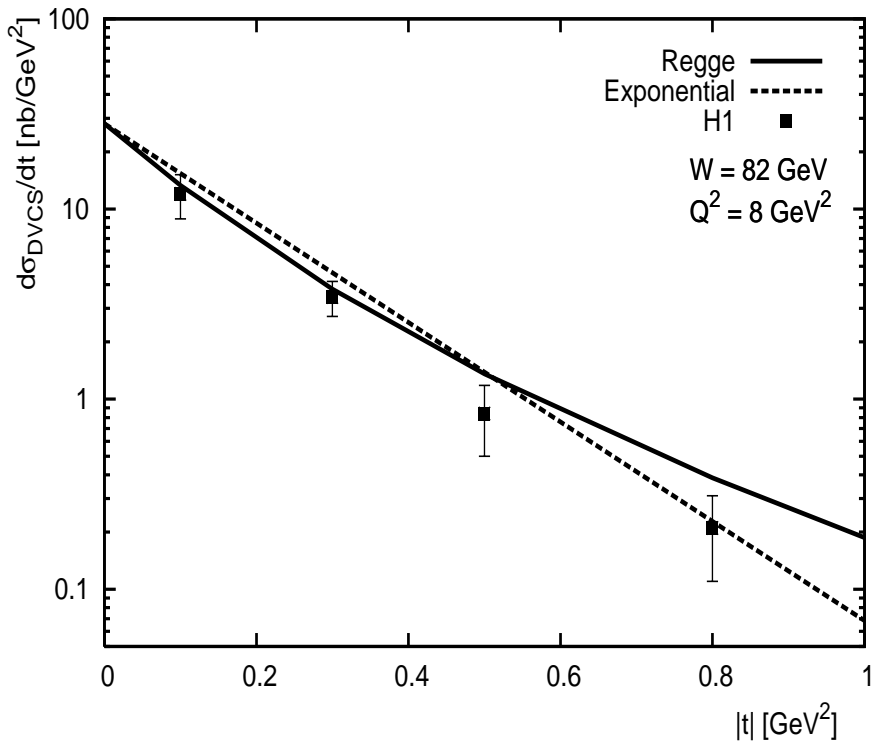,width=12cm,height=10cm}
\caption{The differential DVCS cross section as a function of $|t|$.
The dual parameterization predictions (solid and dashed curves, see details in
the caption to Fig.~\ref{fig:q2vadim}) are compared to the H1 
data~\protect\cite{Aktas:2005ty}.
}
\label{fig:tdepvadim}
\end{center}
\end{figure}

In addition to the $t$-integrated DVCS cross section, for the first time
the H1 reported the differential
 DVCS cross section~~\cite{Aktas:2005ty}. The dual parameterization
 predictions for $d \sigma_{{\rm DVCS}}/dt$ as a function of $t$ are compared to the 
H1 data in Fig.~\ref{fig:tdepvadim}. The theoretical predictions are made using 
Eq.~(\ref{eq:sigmadvcs}) without the integration over $t$
\begin{equation}
\frac{d \sigma_{{\rm DVCS}}(x_B,t,Q^2)}{dt}=\frac{\pi \alpha_{{\rm em}}^2 x_B^2}{Q^4 \sqrt{1+ 4m_N^2 x_B^2/Q^2}} |{\cal A}_{{\rm DVCS}}(\xi,t,Q^2)|^2 \,.
\label{eq:sigmadvcs_tdep}
\end{equation}
As one can see from Fig.~\ref{fig:tdepvadim}, for $|t| < 0.5$ GeV$^2$ both used models
of the $t$-dependence give rather similar predictions and describe well the data. However, 
for $|t| > 0.5$ GeV$^2$, the exponential model corresponds to a steeper decrease of
$d \sigma_{{\rm DVCS}}/dt$ with increasing $|t|$ and allows to describe very well the highest
$|t|=0.8$ GeV$^2$ data point. The experimental errors on $d \sigma_{{\rm DVCS}}/dt$ are small
enough to conclude that that the Regge-motivated model of the $t$-dependence of 
GPDs~(\ref{eq:regge2}) seems to be disfavored by the large-$|t|$ H1 data~\cite{Aktas:2005ty}.

It is important to appreciate that within the framework of the dual
parameterization of GPDs, the HERA data on DVCS cross section were described
so well without any 
free parameters: We used only the forward parton distributions
for the input.
 To be more precise, the parameters $\alpha^{\prime}$ and $\alpha^{\prime}_g$ 
were adjusted to reproduce only the normalization of the DVCS cross section:
The $Q^2$, $W$ and $t$-dependencies of the cross section were then predicted
without any further adjustments.
 The calculation 
using the factorized exponential  $t$-dependence contained no free adjustable
parameters since the slope $B(\mu)$~(\ref{eq:B}) has been experimentally 
measured~\cite{Aktas:2005ty} and, hence, it could not be varied.

The DVCS cross section can be also described using other models of GPDs. 
Within the framework of the double distribution model,
the H1 and ZEUS data on $\sigma_{{\rm DVCS}}$ were successfully described
with the asymptotic ansatz for $h(\beta,\alpha)$ in 
Eq.~(\ref{eq:radyushkin}), which corresponds to the $\xi$-independent input,
$H^i_{{\rm DD}}(x,\xi,\mu_0^2)=(q^i(x,\mu_0^2)+\bar{q}^i(x,\mu_0^2))/2$~\cite{Belitsky:2001ns,Freund:2001hd}.
The observation that the DVCS cross section at high-energies (small $x_B$) and
at high $Q^2$ can be described by the GPDs, whose shape at the low input scale
$\mu^2_0$ is $\xi$-independent, can be qualitatively 
explained as follows.
Under QCD evolution, a GPD at a given small $x$ and large $\mu^2$ is obtained
using the GPD at the low input scale $\mu_0^2$, which is probed for
$x_0 \gg x$. Therefore, the small external parameter $\xi$ can be neglected in the
input GPDs~\cite{Frankfurt:1997ha}.


Other theoretical approaches, which enable one to successfully describe
the HERA data on $\sigma_{{\rm DVCS}}$, include the dipole formalism~\cite{Favart:2003cu,Kowalski:2006hc} and the formalism based on the
conformal moments of the GPDs~\cite{Muller:2006pm}.

\section{DVCS asymmetries}
\label{eq:asymmetry}

Complete expressions for various DVCS asymmetries are well-known~\cite{Belitsky:2001ns}.
In this work, we consider the  beam-spin $A_{LU}$, beam-charge $A_C$ and 
transversely-polarized target $A_{UT}$ asymmetries.
The first two asymmetries are predominantly sensitive to the GPD $H$, while the 
$A_{UT}$ asymmetry is sensitive to both GPDs $H$ and $E$.
Since most of the data on these DVCS asymmetries have come from the HERMES collaboration
at DESY~\cite{Airapetian:2001yk,Seitz:2002si,Ellinghaus:2002bq,Ellinghaus:2004dw,Krauss:2005ja,Ye:2005pf,Ye:2006gz,h:2006zr},
 we will predominantly make numerical predictions for the above asymmetries using the
dual parameterization of GPDs  $H$ and $E$ in the HERMES kinematics.
In addition, predictions for the Jefferson Lab kinematics will be also presented.

\subsection{Beam-spin asymmetry}

Using results of Ref.~\cite{Belitsky:2001ns}, one obtains the 
following approximate expression for the $\sin \phi$-moment of the beam-charge asymmetry
\begin{equation}
A_{LU}^{\sin \phi} \approx +\left(\frac{x_B}{y}\right)\,8\,K\,y\,(2-y)(1+\epsilon^2)^2\,
 \frac{\left[F_1(t) Im \,{\cal H}(\xi,t,Q^2)+\frac{|t|}{4 m_N^2} F_2(t) Im\, {\cal E}(\xi,t,Q^2)\right]}{c_{0,{\rm unp}}^{{\rm BH}}} \,,
\label{eq:alu}
\end{equation}
where $\epsilon=x_B m_N/Q$; the kinematic suppression factor $K$ and 
the leading harmonic of the Bethe-Heitler amplitude squared $c_{0,{\rm unp}}^{{\rm BH}}$ are given in~\cite{Belitsky:2001ns}; $F_1$ and $F_2$ are the Dirac and Pauli proton form factors, 
see e.g.~\cite{Belitsky:2001ns};  ${\cal H}$ and  ${\cal E}$ are defined
by Eq.~(\ref{eq:Hcal}). Equation~(\ref{eq:alu}) is approximate because we have neglected
subleading harmonics (proportional to $c_{1,{\rm unp}}^{{\rm BH}}$ and $c_{2,{\rm unp}}^{{\rm BH}}$)
in the expansion of the Bethe-Heitler amplitude squared and the DVCS amplitude squared in the
denominator of  Eq.~(\ref{eq:alu}). 

Note that we have introduced an additional minus sign in order to take into account
the sign difference between our notation for ${\cal H}$ and the notation of 
Ref.~\cite{Belitsky:2001ns}. Therefore, in Eq.~(\ref{eq:alu}), 
the plus sign corresponds to the 
positively charged lepton beam. Since in our notation $Im \, {\cal H} < 0$ in
the bulk of the considered kinematics,
$A_{LU}^{\sin \phi}$ in the HERMES kinematics is negative.

One should point out that we use the reference frame of Ref.~\cite{Belitsky:2001ns}, which 
differs from the frame used by the HERMES collaboration by the direction of the $z$-axis
(the Trento sign conventions~\cite{Bacchetta:2004jz}).
This means that $\phi_{{\rm HERMES}}=\pi-\phi$~\cite{Belitsky:2001ns,Ellinghaus:2005uc}.
Obviously, for the $\sin \phi$-moment of the beam-spin asymmetry, this difference in the
notations is irrelevant.

Using the dual parameterization of GPD discussed in Sect.~\ref{sec:model} and 
substituting it in
Eq.~(\ref{eq:alu}), we obtain the following range of predictions at the 
average kinematic point of the HERMES measurement, $\langle x_B \rangle=0.11$, $\langle Q^2 \rangle=2.6$
GeV$^2$ and $\langle t \rangle=-0.27$ GeV$^2$~\cite{Airapetian:2001yk},
\begin{eqnarray}
&&A_{LU}^{\sin \phi}=-0.22 \ldots -0.24 \,,\quad {\rm exponential}\ t-{\rm dependence} \,, \nonumber\\
&&A_{LU}^{\sin \phi}=-0.27 \ldots -0.29 \,, \quad {\rm Regge}\ t-{\rm dependence} \,. 
\label{eq:alu_prediction}
\end{eqnarray}
The smaller absolute values of $A_{LU}^{\sin \phi}$ correspond to the calculation with
$J_u=J_d=0$; the larger absolute values of $A_{LU}^{\sin \phi}$ correspond to the 
calculation with $J_u=0.3$ and $J_d=0$ (the variation of $J_d$ has no noticeable affect).
As expected, the small range of predictions (for a given model of the $t$-dependence)
indicates small sensitivity of $A_{LU}^{\sin \phi}$ to the GPD $E$.

Our theoretical calculations compare very well to the HERMES measurement~\cite{Airapetian:2001yk}
\begin{eqnarray}
A_{LU}^{\sin \phi}=-0.23 \pm 0.04 \pm 0.03 \,.
\end{eqnarray}

\begin{figure}[h]
\begin{center}
\epsfig{file=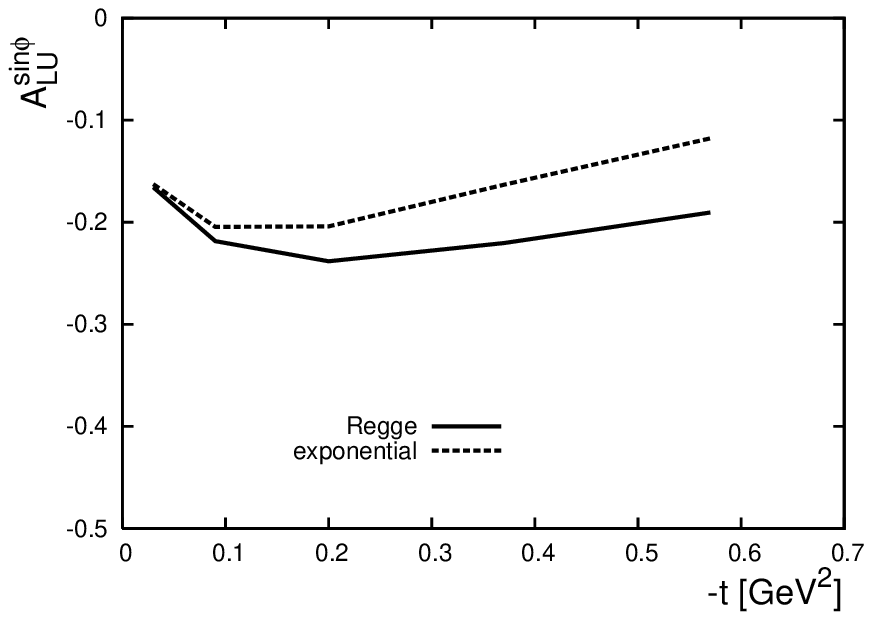,width=8cm,height=8cm}
\epsfig{file=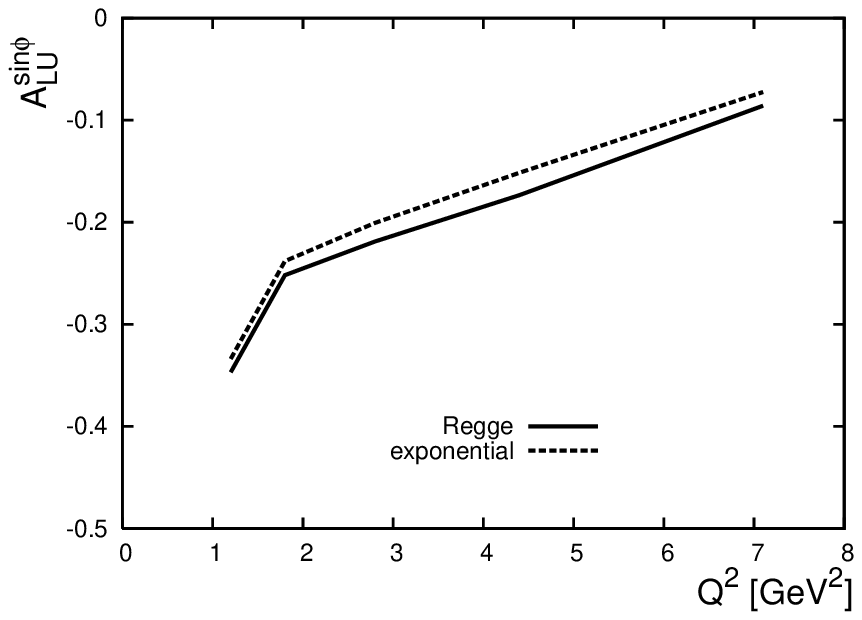,width=8cm,height=8cm}
\epsfig{file=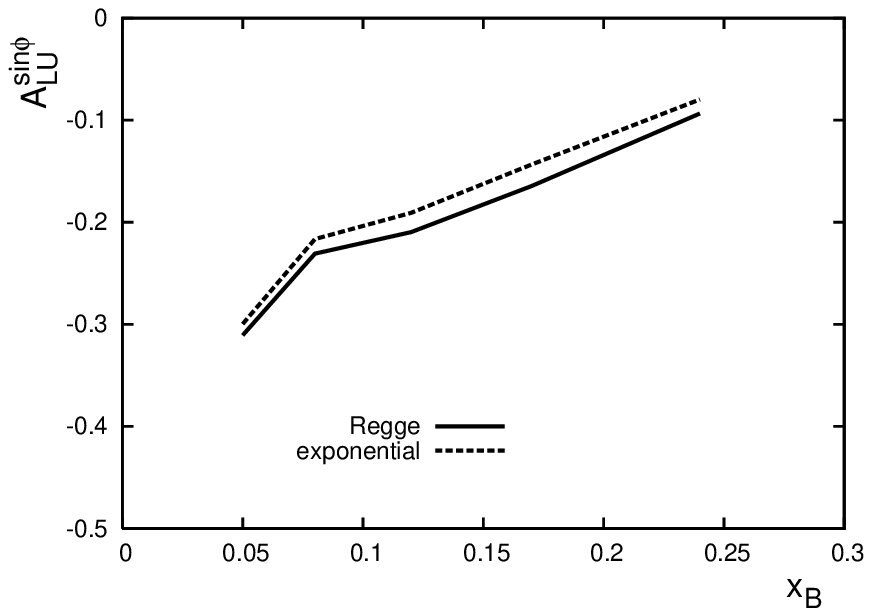,width=8cm,height=8cm}
\caption{$A_{LU}^{\sin \phi}$ as a function of  $t$, $x_B$ and $Q^2$. The 
dual parameterization predictions with two models of the $t$-dependence
in the HERMES kinematics~\protect\cite{Ellinghaus:2004dw}.}
\label{fig:beamspin}
\end{center}
\end{figure}

In addition to the average HERMES kinematics, we studied the dependence
of $A_{LU}^{\sin \phi}$ on $t$, $x_B$ and $Q^2$ bin by
 bin~\cite{Ellinghaus:2004dw,Ellinghaus:2005uc}.
Figure~\ref{fig:beamspin} summarizes our predictions, which are
made using $J_u=J_d=0$.

As can be seen from Figure~\ref{fig:beamspin}, only $A_{LU}^{\sin \phi}$
as a function of $|t|$ can be helpful in distinguishing between the exponential
and Regge models of the $t$-dependence, provided the experimental uncertainties
are sufficiently small.


We also make predictions for the beam-spin asymmetry in the CLAS kinematics.
For the 2001 average kinematic point of the CLAS kinematics~\cite {Stepanyan:2001sm},
$E=4.25$ GeV, $\langle Q^2 \rangle =1.25$ GeV$^2$, $\langle x_B \rangle =0.19$ and
$\langle t \rangle=-0.19$ GeV$^2$, our predictions compare very well to the 
experimental value,
\begin{eqnarray}
&&A_{LU}^{\sin \phi}=0.15 \ldots 0.17 \,,\quad {\rm exponential}\ t-{\rm dependence} \,, \nonumber\\
&&A_{LU}^{\sin \phi}=0.18 \ldots 0.20 \,, \quad {\rm Regge}\ t-{\rm dependence} \,,
 \nonumber\\
&&A_{LU}^{\sin \phi}=0.202 \pm 0.028 \,, \quad {\rm CLAS}~[58] \,.
\label{eq:alu_prediction_clas}
\end{eqnarray}
The lower values of the theoretical predictions correspond to
$J_u=J_d=0$; the larger values correspond to $J_u=0.3$ and $J_d=0$.
Note the sign change in  $A_{LU}^{\sin \phi}$ when going from the positron 
beam (HERMES) to the electron beam (CLAS).

\begin{figure}[t]
\begin{center}
\epsfig{file=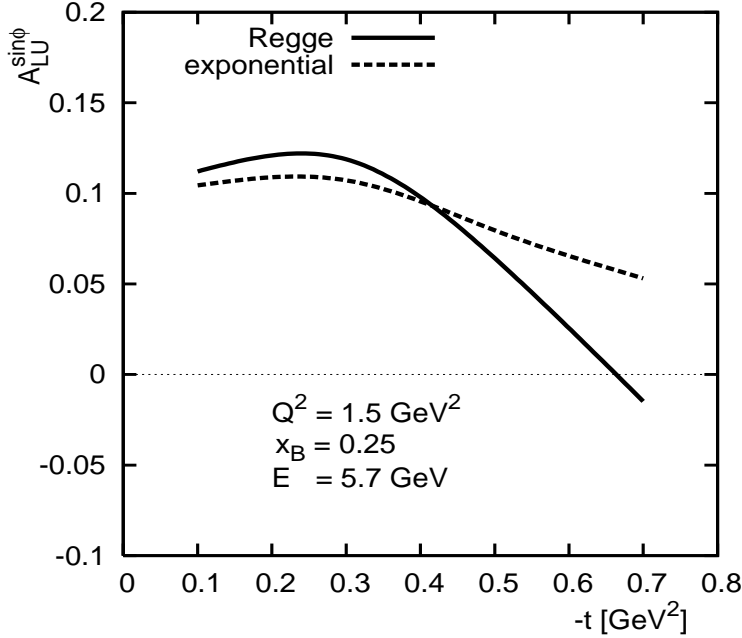,width=10cm,height=9cm}
\caption{$A_{LU}^{\sin \phi}$ as a function of  $t$ in the CLAS kinematics.
The two curves represent calculations with two models of the $t$-dependence (see the text).}
\label{fig:beamspin_clas}
\end{center}
\end{figure}

Recently, CLAS performed dedicated measurements of DVCS and, in particular,
of the beam-spin asymmetry with higher energies of the lepton beam
and with much wider kinematic coverage in $Q^2$, $x_B$ and $t$.
Figure~\ref{fig:beamspin_clas} presents our predictions for the $t$-dependence of
$A_{LU}^{\sin \phi}$ at $E=5.7$ GeV, $Q^2 =1.5$ GeV$^2$ and $x_B=0.25$.

It is important to point out that, by construction,
the minimal model of the dual parameterization of GPDs is designed for 
small values of $x_B$, $x_B \leq 0.2$. The increase of $x_B$ from 
$x_B \approx 0.2$ (HERMES) to $x_B \approx 0.3$ (current CLAS) 
leads to the increasing role
of the generating functions $Q_2^i$ and $R_2^i$ (the latter plays role
at large $|t|$, $|t| > 0.5$ GeV$^2$), which results in a significant 
model-dependence of our results. Therefore, our predictions in
Fig.~\ref{fig:beamspin_clas} should be taken as semi-quantitative.
Strictly speaking, in order to make 
quantitative predictions for the current CLAS kinematics, one has to extend
the minimal model by including higher generating functions 
$Q^i_k$ and $R^i_k$ with $k \geq 4$.

\subsection{Beam-charge asymmetry}

Next we turn to the beam-charge asymmetry, $A_C$. Using the results of 
Ref.~\cite{Belitsky:2001ns}, the approximate expression for $A_C$ reads
[we have neglected the same terms in the denominator as in Eq.~(\ref{eq:alu})]
\begin{equation}
A_C(\phi) \approx \left(\frac{x_B}{y}\right)\,\left(1+\epsilon^2\right)^2\,\frac{c_{0,{\rm unp}}^{{\cal I}}+
c_{1,{\rm unp}}^{{\cal I}}\cos \phi}{c_{0,{\rm unp}}^{{\rm BH}}} \,,
\label{eq:ac}
\end{equation}
where $c_{0,{\rm unp}}^{{\cal I}}$ and $c_{1,{\rm unp}}^{{\cal I}}$ are given in 
Ref.~\cite{Belitsky:2001ns}.
While $c_{0,{\rm unp}}^{{\cal I}}$ is smaller than $c_{1,{\rm unp}}^{{\cal I}}$, it is not
negligibly small. In this work, we shall concentrate on the larger contribution to
$A_C$ proportional to $c_{1,{\rm unp}}^{{\cal I}}$, which can be singled out by considering 
the $\cos \phi$-moment of $A_C$
\begin{equation}
A_{C}^{\cos \phi} \approx - \left(\frac{x_B}{y}\right)\,8\,K\,(2-2y+y^2)\,\left(1+\epsilon^2\right)^2\,
 \frac{\left[F_1(t)\, Re\, {\cal H}(\xi,t,Q^2)+\frac{|t|}{4 m_N^2} F_2(t)\, Re \,{\cal E}(\xi,t,Q^2)\right]}{c_{0,{\rm unp}}^{{\rm BH}}} \,.
\label{eq:ac2}
\end{equation}
As discussed above, $\phi_{{\rm HERMES}}=\pi-\phi$. Therefore,
$A_{C}^{\cos \phi_{{\rm HERMES}}}=-A_{C}^{\cos \phi}$.
Since in the considered kinematics $Re\, {\cal H} >0$, we obtain
$A_{C}^{\cos \phi_{{\rm HERMES}}} > 0$.
Until the end of this subsection, we shall imply $\phi_{{\rm HERMES}}$, but we
will use $\phi$ for brevity.

The range of predictions for $A_{C}^{\cos \phi}$ using the dual parameterization
of GPDs can be compared to the HERMES measurements. 
For the 2002 HERMES  average kinematic point, $\langle x_B \rangle=0.12$,
 $\langle Q^2 \rangle=2.8$
GeV$^2$ and $\langle t \rangle=-0.27$ GeV$^2$~\cite{Ellinghaus:2002bq},
we obtain
\begin{eqnarray}
&&A_{C}^{\cos \phi}=0.01 \ldots 0.03 \,,\quad {\rm exponential}\ t-{\rm dependence} \,, \nonumber\\
&&A_{C}^{\cos \phi}=0.19 \ldots 0.23 \,, \quad {\rm Regge}\ t-{\rm dependence} \,,
\nonumber\\
&&A_{C}^{\cos \phi}=0.11 \pm 0.04 \pm 0.03 \,, \quad {\rm HERMES}~[50] \,.
\label{eq:ac_prediction}
\end{eqnarray}
The lower values of $A_{C}^{\cos \phi}$ correspond to the calculation with
$J_u=J_d=0$; the larger values correspond to $J_u=0.3$ and $J_d=0$.

For the very recent 2006 HERMES average kinematic point, $\langle x_B \rangle=0.10$,
 $\langle Q^2 \rangle=2.5$
GeV$^2$ and $\langle t \rangle=-0.12$ GeV$^2$~\cite{h:2006zr},
we obtain
\begin{eqnarray}
&&A_{C}^{\cos \phi}=0.013 \ldots 0.022 \,,\quad {\rm exponential}\ t-{\rm dependence} \,, \nonumber\\
&&A_{C}^{\cos \phi}=0.080 \ldots 0.092 \,, \quad {\rm Regge}\ t-{\rm dependence} \,,
\nonumber\\
&&A_{C}^{\cos \phi}=0.063 \pm 0.029 \pm 0.026 \,, \quad {\rm 
HERMES}~[55]\,.
\label{eq:ac_prediction_new}
\end{eqnarray}

The following two features of Eqs.~(\ref{eq:ac_prediction})
and  (\ref{eq:ac_prediction_new}) deserve a discussion.
First, the exponential model of the $t$-dependence predicts 
the values of $A_{C}^{\cos \phi}$, which
are much smaller than those calculated with the Regge model of the 
$t$-dependence. The reasons for this are the non-trivial $t$-dependence of the
real part of ${\cal H}$ (more precisely, the non-trivial cancellation
between two contributions to the real part of ${\cal H}$)
 and the large values of $|t|$ involved.

Second, predictions with the exponential model of the $t$-dependence are much more
sensitive to the fraction of proton spin carried by the quarks, $J^i$.

In addition to the average kinematic point, the recent HERMES 
analysis~\cite{h:2006zr} presented $A_{C}^{\cos \phi}$ as a function of $t$.
\begin{figure}[h]
\begin{center}
\epsfig{file=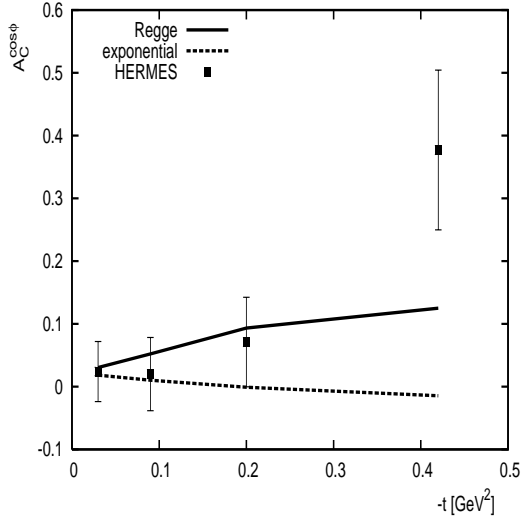,width=7cm,height=7cm}
\caption{$A_{C}^{\cos \phi}$ as a function of  $t$. The 
dual parameterization predictions with two models of the $t$-dependence are
compared to the HERMES data~\protect\cite{h:2006zr}.}
\label{fig:beamcharge_recent}
\end{center}
\end{figure}
\begin{figure}[h]
\begin{center}
\epsfig{file=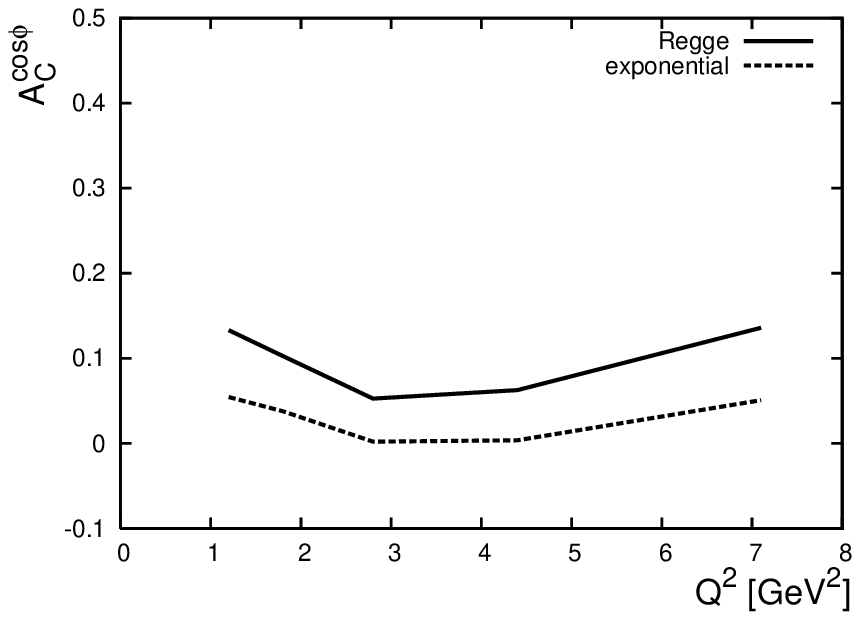,width=7cm,height=7cm}
\epsfig{file=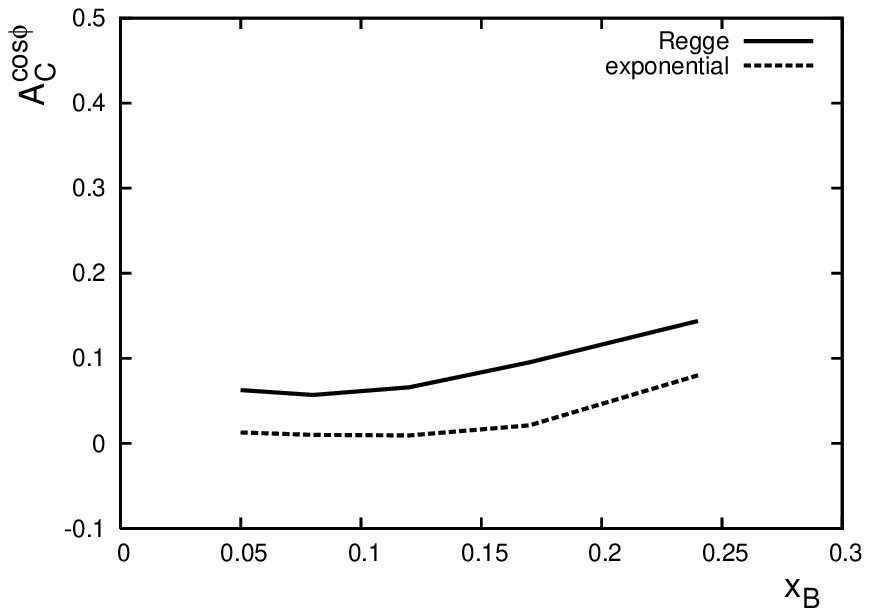,width=7cm,height=7cm}
\caption{$A_{C}^{\cos \phi}$ as a function of $Q^2$ and $x_B$.
The 
dual parameterization predictions with two models of the $t$-dependence
in the HERMES kinematics~\protect\cite{Ellinghaus:2004dw}.}
\label{fig:beamcharge}
\end{center}
\end{figure}
Figure~\ref{fig:beamcharge_recent} presents the comparison of our theoretical 
predictions to the data. It can be seen from Fig.~\ref{fig:beamcharge_recent}
that the Regge model describes the data points well for $|t| < 0.2$ GeV$^2$
and underestimates the asymmetry for the larger $|t|  \approx 0.4$ GeV$^2$. 
The exponential model of the $t$-dependence dramatically 
fails to describe the rise of $A_{C}^{\cos \phi}$ with increasing  $|t|$.
Therefore, on the basis of the comparison of our theoretical predictions to the 
$t$-dependence of the $\cos \phi$-moment of the beam-charge asymmetry, we conclude that
the non-factorized Regge model of the $t$-dependence of GPDs is preferred over the
factorized exponential model.

In addition to the $t$-dependence of $A_{C}^{\cos \phi}$, we make 
predictions for the $Q^2$ and $x_B$-dependence of $A_{C}^{\cos \phi}$ in the 
HERMES kinematics~\cite{Ellinghaus:2004dw,Ellinghaus:2005uc}
in Fig.~\ref{fig:beamcharge}. 
Our theoretical predictions are made using $J_u=J_d=0$.

\subsection{Transversely polarized target asymmetry} 

The DVCS asymmetry with the unpolarized beam and the transversely polarized target,
$A_{UT}$,
is sensitive to all four GPDs of the nucleon~\cite{Belitsky:2001ns}.
Since we are concerned with the GPDs $H$ and $E$, we
shall consider the $\sin \varphi \cos \phi$-moment of $A_{UT}$, where the angle
$\varphi$ is the angle between the vector of the target polarization and the hadron scattering
plane in the notation of Ref.~\cite{Belitsky:2001ns}. The main interest in considering this
DVCS observable is that it is sensitive to the GPD $E$ and, hence, to the 
fraction of the proton total angular momentum carried by quarks, $J^i$.

According to the Trento sign convention~\cite{Bacchetta:2004jz}, it
is recommended to use different angles, which are used e.g.~in the HERMES analysis:
$\phi_{{\rm HERMES}}=\pi-\phi$ and 
$\phi_{{\rm S\,,HERMES}}-\phi_{{\rm HERMES}}=\pi+\varphi$~\cite{Ellinghaus:2005uc}.
Obviously, $A_{UT}^{\sin \varphi \cos \phi}=A_{UT}^{\sin (\phi_{{\rm S\,,HERMES}}-\phi_{{\rm HERMES}})
\cos \phi_{{\rm HERMES}}}$.

Using the results~\cite{Belitsky:2001ns}, the approximate expression for $A_{UT}^{\sin \varphi \cos \phi}$ reads
\begin{eqnarray}
&&A_{UT}^{\sin \varphi \cos \phi}  \approx   \left(\frac{x_B}{y}\right) \left(1+\epsilon^2\right)^2 
\frac{1}{c_{0,{\rm unp}}^{{\rm BH}}} \frac{8\, m_N \sqrt{1-y}}{Q} \left(2-2\,y+y^2\right)
\nonumber\\
&\times& \Big[\frac{1}{2-x_B} \left(x_B^2 F_1(t)-(1-x_B) \frac{t}{m_N^2}F_2(t)\right) Im\, {\cal H}(\xi,t,Q^2)
\nonumber\\
&+&\left(\frac{x_B^2}{2-x_B}F_1(t) + \frac{t}{4\,m_N^2}\left((2-x_B)F_1(t)+\frac{x_B^2}{2-x_B} F_2(t) \right)\right) Im\, {\cal E}(\xi,t,Q^2)\Big] \,.
\label{eq:A_UT}
\end{eqnarray}
Note that, similarly to the above considered $A_{LU}^{\sin \phi}$ and  $A_{C}^{\cos \phi}$,
we have introduced an additional minus sign to compensate the sign difference between our
definition of ${\cal H}$ and ${\cal E}$ and the notation of Ref.~\cite{Belitsky:2001ns}.

\begin{figure}[t]
\begin{center}
\epsfig{file=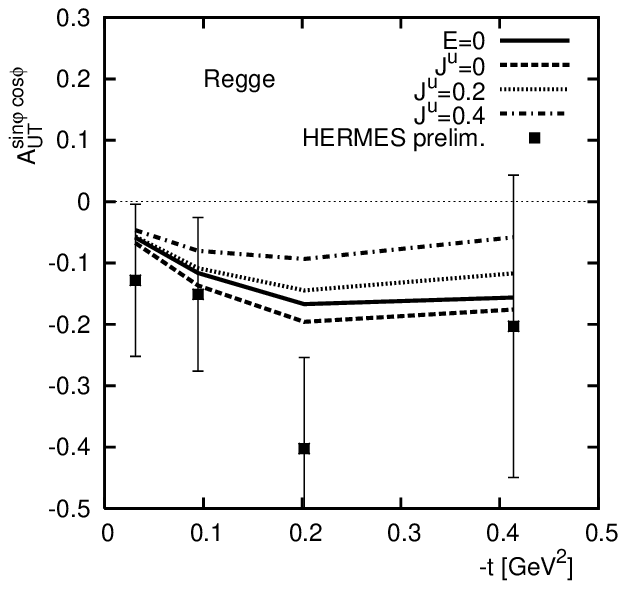,width=6cm,height=5cm}
\epsfig{file=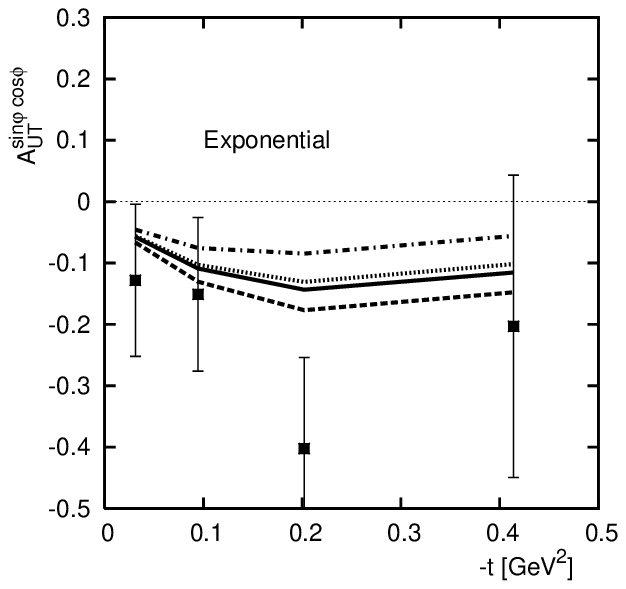,width=6cm,height=5cm}
\epsfig{file=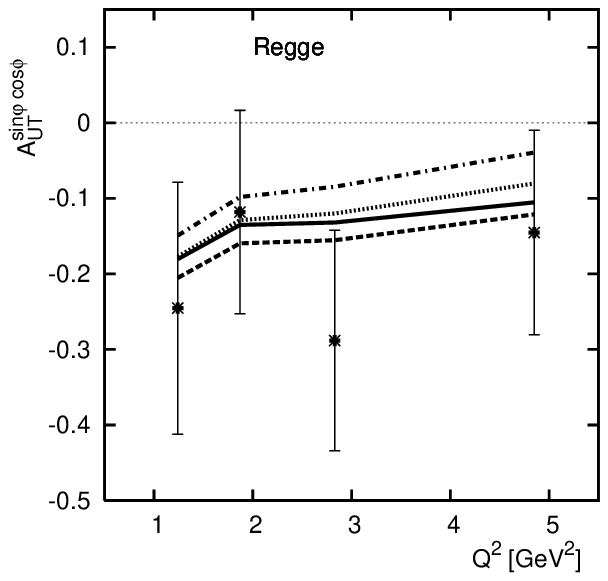,width=6cm,height=5cm}
\epsfig{file=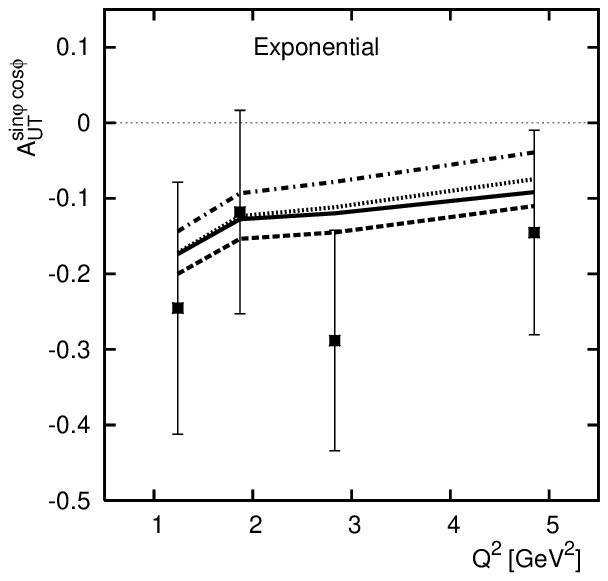,width=6cm,height=5cm}
\epsfig{file=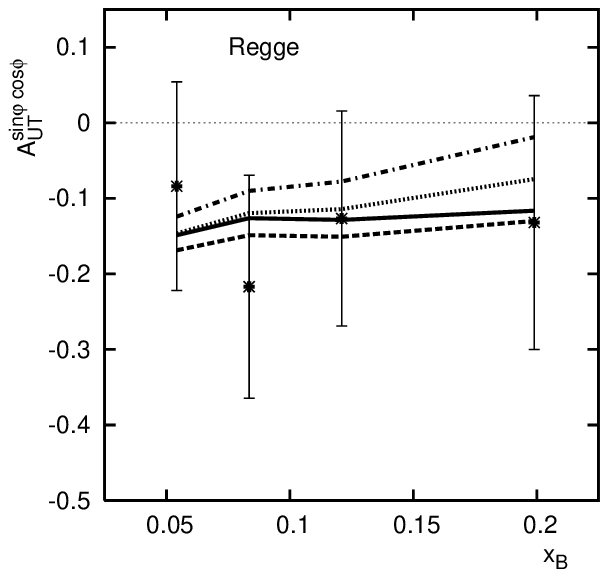,width=6cm,height=5cm}
\epsfig{file=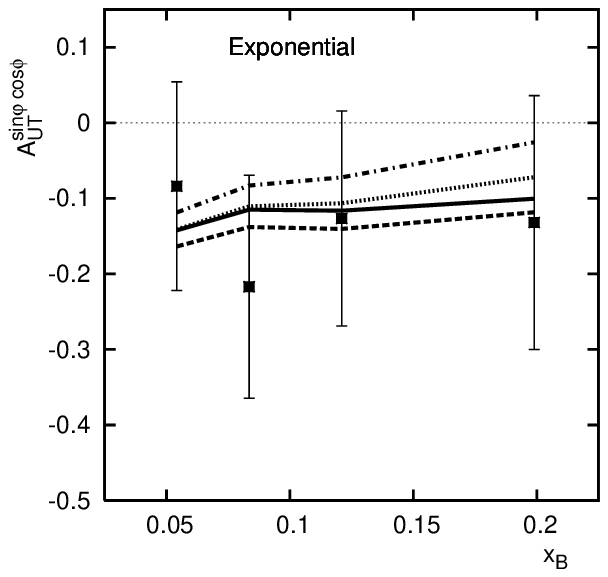,width=6cm,height=5cm}
\caption{$A_{UT}^{\sin \varphi \cos \phi}$ as a function of  $t$, $x_B$ and $Q^2$. The 
dual parameterization predictions with two models of the $t$-dependence are
compared to the HERMES preliminary data~\protect\cite{Ye:2005pf}.
The error bars correspond to the statistical and systematic 
uncertainties added in quadrature. The systematic uncertainty does not
include the effect of the HERMES acceptance.}
\label{fig:tr}
\end{center}
\end{figure}

The theoretical predictions for the $t$, $Q^2$ and $x_B$-dependence of  $A_{UT}^{\sin \varphi \cos \phi}$
using the dual parameterization of GPDs and  Eq.~(\ref{eq:A_UT}) 
are compared to the preliminary HERMES data~\cite{Ye:2005pf} in Fig.~\ref{fig:tr}.
Note that the shown error bars correspond to the statistical and systematic 
uncertainties added in quadrature and that the systematic uncertainty does not
include the effect of the HERMES acceptance. 

The plots for the $t$ and $Q^2$-dependence appear to be most informative.
As can be seen from the upper and middle panels of Fig.~\ref{fig:tr}, our theoretical calculations reproduce the data fairly well,
except for one point. 
Judging by the central experimental values, one concludes that the data seems to 
prefer the scenario with $J_u=J_d=0$. 

Also, both models of the $t$-dependence give
rather close results. Therefore, in view of large experimental uncertainties, 
it is impossible to differentiate between the Regge and exponential models
of the $t$-dependence of the GPDs using the current preliminary HERMES data on
$A_{UT}^{\sin \varphi \cos \phi}$~\cite{Ye:2005pf}.

In addition, we compare our predictions for $A_{UT}^{\sin \varphi \cos \phi}$
to the HERMES measurement at the average kinematic point,
$\langle x_B \rangle =0.095$, $\langle Q^2 \rangle=2.5$ GeV$^2$
and $\langle -t \rangle=0.12$ GeV$^2$~\cite{Ye:2006gz},
\begin{eqnarray}
&& A_{UT}^{\sin \varphi \cos \phi}=-0.14 \ldots -0.10 \,, \quad {\rm exponential}\ t-{\rm dependence} \,, \nonumber\\
&&A_{UT}^{\sin \varphi \cos \phi}=-0.15 \ldots -0.10 \,, \quad {\rm Regge}\ t-{\rm dependence} \,,
\nonumber\\
&&A_{C}^{\cos \phi}=-0.149 \pm 0.058 \pm 0.033 \,, \quad {\rm 
prelim.\ HERMES}~[54]\,.
\label{eq:A_UT_average}
\end{eqnarray}
The lower absolute values of $A_{UT}^{\sin \varphi \cos \phi}$ correspond to the calculation with
$J_u=J_d=0$; the larger values correspond to $J_u=0.3$ and $J_d=0$.
As can be seen from Eq.~(\ref{eq:A_UT_average}), the agreement between our 
predictions and the experimental value is very good.
The central experimental value prefers $J_u=J_d=0$.

Predictions for $A_{UT}^{\sin \varphi \cos \phi}$ were also made within the framework 
of the double distribution model of GPDs~\cite{Ellinghaus:2005uc}. 
The comparison of the theretical predictions~\cite{Ellinghaus:2005uc} to the HERMES 
data~\cite{Ye:2005pf} does not allow one to make a definite conclusion about the fraction
of the proton total angular momentum carried by the $u$-quark: The data on the $t$-dependence somewhat
prefers $J_u=0$ and $J_u=0.2$, while the data on $x_B$ and $Q^2$-dependence
prefer $J_u=0.2$ and $J_u=0.4$.

In conclusion of this short discussion of the fraction of the total angular momentum
of the proton carried by quarks, we would like to mention the 
lattice results of the QCDSF collaboration:
$J^u=0.37 \pm 0.06$ and $J^d=-0.04 \pm 0.04$~\cite{Gockeler:2003jf}.
Our analysis presented in this section indicates that the value of $J^u$ should
be significantly smaller.

\section{Conclusions and discussion}
\label{sec:conclusions}

In this work, we considered the new leading order (LO) dual parameterization
of GPDs introduced by Shuvaev and Polyakov~\cite{Polyakov:2002wz}. 
The advantages of the dual parameterization include simple (forward) QCD evolution 
of resulting GPDs and a
simple expression for the LO DVCS amplitude. 
We extended the work
by Guzey and Polyakov~\cite{Guzey:2005ec} and formulated the minimal model of
the dual parameterization of GPDs $H^i$ and $E^i$, which enables one to relate the 
GPDs to the fairly well-known quantities. In particular, apart from the $t$-dependence,
the GPDs $H^i$ can be formulated in terms of  the forward quark distributions
$q^i$ and Gegenbauer moments of the $D$-term.
The GPDs $E^i$ can be formulated in terms of the unknown forward limit of the GPDs $E^i$
and, again, the Gegenbauer moments of the $D$-term.
The price to pay for the simplicity of our dual model is that the model is 
designed for not too large $x_B$, $x_B \leq 0.2$. 
Within the considered model, the $t$-dependence of GPDs has to be modelled separately.
We considered two different models of the $t$-dependence: The factorized exponential
model and the non-factorized Regge-motivated model.

We compared predictions of our model to all available data on DVCS cross section
and asymmetries. The $Q^2$, $W$ and $t$-dependence of the DVCS cross section at high energies (small $\xi$) measured by the H1 and ZEUS collaborations were successfully described 
by both models of the $t$-dependence. It should be stressed that our predictions for 
$\sigma_{{\rm DVCS}}$ are virtually model-independent: Only the normalization of the 
cross section at one kinematic point was fitted by appropriately choosing the 
effective slope parameters in the Regge-motivated model of the $t$-dependence.

Turning to the beam-spin DVCS asymmetry, $A_{LU}$, we successfully described both HERMES and
CLAS data on $A_{LU}^{\sin \phi}$ at their respective average kinematic points. We also made predictions
for the $t$, $Q^2$ and $x_B$-dependence of $A_{LU}^{\sin \phi}$ in the HERMES kinematics bin-by-bin 
and for the $t$-dependence of $A_{LU}^{\sin \phi}$ in the CLAS kinematics
 with $E=5.7$ GeV.
We observed that only the $t$-dependence of $A_{LU}^{\sin \phi}$ has a chance to distinguish between the
two considered models of the $t$-dependence.

We found that within our framework, the beam-charge asymmetry $A_{C}$ is the only 
considered observable which distinguishes between the Regge-motivated and exponential 
models of the $t$-dependence of GPDs. While the Regge-motivated model provides a reasonable 
description of $A_{C}^{\cos \phi}$ in the average HERMES kinematics and of
the $t$-dependence of
$A_{C}^{\cos \phi}$ measured at HERMES, the exponential model of the $t$-dependence
fails dramatically.

We also compared our predictions to the HERMES measurement of the DVCS asymmetry 
measured with the unpolarized beam and the transversely-polarized target,
$A_{UT}$. We obtained a fairly good description of the preliminary HERMES data
on $t$, $Q^2$ and $x_B$-dependence of $A_{UT}^{\sin \varphi \cos \phi}$
using both models of the $t$-dependence. While the experimental uncertainties are
large, the data still seems to indicate that, within our model, $J_u=J_d \approx 0$,
i.e.~that
the $u$ and $d$ quarks carry only a small fraction of the proton total
angular momentum.

All comparisons to the experimental values presented in this work were
done taking the minimal version of the dual parameterization of GPDs at its face
value. We did not take into account such potentially important effects as
next-to-leading order corrections and higher twist effects.
Their role in the context of the dual parameterization of GPDs is a subject
of a separate analysis.

In conclusion, the dual parameterization of GPDs presents a new model of GPDs, 
which, with a small number of model-dependent inputs, allows for a uniform
description of all available data on DVCS cross section and asymmetries.

\acknowledgements

The authors would like to thank Z. Ye and D. Hasch for providing us with the
preliminary HERMES data on $A_{UT}^{\sin \varphi \cos \phi}$. V.G.~would also like to thank
D.~M\"uller for useful comments.

\end{document}